\documentclass[aps,prb,twocolumn,showpacs,amssymb]{revtex4-1}
\usepackage{amssymb,amsmath,amsfonts}
\usepackage{graphicx,subfigure,epsfig}
\usepackage{natbib}
\usepackage{dcolumn}% Align table columns on decimal point
\usepackage{bm,times}
\usepackage{float,makeidx,longtable}
\usepackage{color}
\usepackage[colorlinks,citecolor=blue,linkcolor=red]{hyperref}

\begin{document}

%
% ------------------------------ Title --------------------------------------- %
%

\title{Synthesizing and Controlling Helical Indirect Exchange Interactions at Nonequilibrium}

% ------------------------------- Authors ------------------------------------ %

\author{Tingting Liu$^1$}
\author{Jie Ren$^2$}
 \email{xonics@tongji.edu.cn}
\author{Peiqing Tong$^{1,3}$}
 \email{pqtong@njnu.edu.cn}

\affiliation{$^1$Department of Physics and Institute of Theoretical Physics, Nanjing Normal
University, Nanjing 210023, P. R. China\\
$^2$Center for Phononics and Thermal Energy Science, China-EU Joint Center for Nanophononics, Shanghai Key Laboratory of Special Artificial Microstructure Materials and Technology, School of Physics Science and Engineering, Tongji University, Shanghai 200092, P. R. China\\
$^3$Jiangsu Key Laboratory for Numerical Simulation of Large Scale Complex Systems, Nanjing Normal University, Nanjing 210023, P. R. China}

\date{\today \\[0.3cm]}

%
% ------------------------------- Abstract ----------------------------------- %
%

\begin{abstract}
We study the nonequilibrium effects of spin and/or electric currents on the helical indirect exchange interactions of local spins that embedded in general open electronic systems. Especially, besides the synthesized anisotropic Heisenberg interactions, we find that the synthetic helical indirect exchange interactions possess two parts: antisymmetric (Dzyaloshinskii-Moriya interaction) and symmetric (Kaplan-Shekhtman-Entin-Wohlman-Aharony interaction), which are all formulated in terms of Keldysh nonequilibrium Green's functions.
The presence of either spin-orbit coupling or spin polarized currents alone is able to synthesize and control the antisymmetric Dzyaloshinskii-Moriya exchange interactions, as the same direction as spin splitting. However, the appearance of symmetric Kaplan-Shekhtman-Entin-Wohlman-Aharony interactions requires both, i.e., the spin-orbit coupling and spin polarized currents with different splitting directions.
Our results show the detailed scheme of controlling the sign, magnitude, and direction of indirect Dzyaloshinskii-Moriya vectors and Kaplan-Shekhtman-Entin-Wohlman-Aharony interactions at nonequilibrium in open quantum devices.
\end{abstract}

%\pacs{03.67.Bg, 03.65.Ud, 61.44.Br, 05.40.Fb}
\maketitle
%
% ----------------------------------- Text ----------------------------------- %
%

\section{Introduction}\label{sec:intoduc}

The antisymmetric helical Dzyaloshinskii-Moriya (DM) interaction was first proposed by Dzyaloshinskii for explaining weak ferromagnetism in antiferromagnetic compounds \cite{ID1958}, and the microscopic basis for this theory was later given by Moriya \cite{T1960}, who extended Anderson superexchange theory \cite{PWA1959} to include the spin-orbit interactions.
The DM interaction plays an essential role for the formation of topologically nontrivial spin structures. For instance, cycloidal spin spirals \cite{MB2007,PF2008,SHP2014}, N\'{e}el type domain walls \cite{AK2003,MHG2008,SM2009,GC2013,SE2013,KR2013}, magnon Hall effect, topological magnon insulators, Skyrmions \cite{AN1989,AN2001,SMB2009,XZ2010,SS2012,AF2013,NN2013,RW2016} and Skyrmion lattices \cite{SH2011,NR2013,NR2015,GC2015,WJ2015,OBJ2016,CML2016,SWL2016}, many of which stimulated interest in fundamental magnetism studies and provided new possibilities for the development of future spintronic devices.
In addition, DM interaction is also very important in the quantum dot quantum computing \cite{NE2001,SC2006}, the quantum phase transitions \cite{RJ2008,ZM2013}, the quantum heat conduction \cite{LW2012}, ferroelectricity in many multiferroics \cite{ss2008,lq2008,si2006} and the properties of entanglement \cite{MK2008,MK2009,BQL2001,SHC2018}.

In the original paper of Moriya \cite{T1960}, it is also found that the antisymmetric DM interaction is also companied by a symmetric helical anisotropy interaction that was one order of magnitude smaller than the DM exchange and thus was neglected.
Nevertheless, Kaplan then found this symmetric helical interaction in the single-band Hubbard model with spin-orbit couplings (SOCs)~\cite{TAK1983} and pointed out its importance. After that, Shekhtman, Entin-Wohlman, and Aharony found that this non-negligible symmetric helical exchange interaction can explain the weak ferromagnetism of $\mathrm{La_2CuO_4}$~\cite{LSO1992}.
Therefore, the symmetric helical exchange interaction was called KSEA interaction for short \cite{JCM2013,AZS1998,AZS1999,ITJ2000,SMS2012,SMG2017,HSK2000,MDL2001,IT2003}. Since then, many clear experimental evidences of the presence of the KSEA interaction came from magnetization measurements, neutron diffraction and inelastic neutron scattering experiments on $\mathrm{Ba_2CuGe_2O_7}$ \cite{AZS1998,AZS1999,ITJ2000,SMS2012,SMG2017}, $\mathrm{Yb_4As_3}$ \cite{HSK2000}, $\mathrm{K_2V_3O_8}$ \cite{MDL2001}, and $\mathrm{La_2CuO_4}$ \cite{IT2003}.

Due to the fact that the helical DM and KSEA interactions are very important, people have tried to synthesize some materials with these helical exchange interactions by artificial material engineering. The strength and the sign of DM and KSEA interactions can be controlled by the material composition, stack order, mixing impurities, and so on.
Alternative effective way to control the magnetic exchange interactions is through the indirect exchange interaction of Ruderman-Kittel-Kasuya-Yoshida (RKKY) mechanism~\cite{Ruderman1954,Kasuya1956,Yosida1957} between molecular magnets.
Fransson-Ren-Zhu~\cite{Ren2014} provided a general framework for the indirect magnetic exchange interactions controlled at nonequilibrium open systems.
The results show that the RKKY interaction presents the Heisenberg interaction regardless of the spin polarization in the molecules, while the nonequilibrium RKKY interaction presents the indirect DM interactions only under spin polarized conditions.
Since then, Fransson and other colleagues %$\emph{et. al}$
found that one can control the molecular spin states and the current flow through the system by the ferro- and antiferro-magnetic switching~\cite{TSJ2016}, the
charge and thermal transport properties of magnetically active paramagnetic molecular dimer~\cite{JD2017}, and the dynamics of a localized molecular spin under the influence of external voltage pulses~\cite{HH2018}. X. Shi $\emph{et. al}$ also found that the DM interactions in a quantum spin system can induce a faster and more efficient quantum state transfer~\cite{XS2017}.

Yet, the indirect KSEA interaction is seldom investigated in literature.
Recently, people \cite{Lee2015,Shiranzaei2017} studied the equilibrium RKKY interaction mediated by the helical edge spin currents in topological insulator \cite{Hasan2010,Qi2011} with Rashba spin-orbit coupling (RSOC) \cite{Rashba1960,Bychkov1984}. With the particular model, they found that in the presence of RSOC, the RKKY interaction induced by the helical edge spin currents contains not only the Heisenberg interaction and the DM interaction, but also a nematic type interaction that is not present in the absence of the RSOC. The nematic type interaction is actually reminiscent of symmetric KSEA interactions.

In this paper, we provide the detailed theoretical scheme to synthesize and control the helical indirect DM and KSEA exchange interactions with nonequilibrium spin and/or electric currents.
The paper is organized as follows. In Section \ref{sec:model} we present the Hamiltonian
of general model and give the derivation of the RKKY interaction in terms of general Keldysh nonequilibrium Green's functions. In Section \ref{sec:result}, the explicit analytic results of the magnetic exchange interactions for two sites system are given and discussed, respectively. Section \ref{sec:summary} is a short summary.

\section{MODELS AND METHODS}\label{sec:model}
\begin{figure}[H]
\centering
\includegraphics[width=0.48\textwidth]{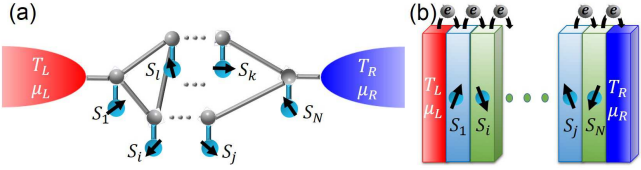}
\vspace{-0.6cm}
\caption{(a) Schematic of localized spins embedded in the sites of an arbitrary electron networks, which finally forms an effective spin network. The sites can be quantum dots, impurities or magnetic molecules. (b) Localized spins are embedded in each layer with the tunneling electron being exchanged among them. Both example systems are connected to external reservoirs, so that the indirect exchange interaction among local spins can be tuned in open quantum systems  at nonequilibrium to form extraordinary spin orders.}\label{fig:1}
\end{figure}

The total system is an open quantum system at nonequilibrium, where a local-spin-electron-hybridized central system (i.e. a molecular or impurity spin-electron coupled network, a multilayer structure where localized spins are embedded in each layer, \emph{et al}) connects with two external electronic reservoirs at different temperatures and chemical potentials (see Fig.~\ref{fig:1}).
The total Hamiltonian is generally described as $H=H_S+\sum_{v=L,R}H_{v}+V_T$, where
$H_{v}=\sum\limits_{{\alpha}k_{v}}(\varepsilon_{{\alpha}k_{v}}-\mu_{v{\alpha}})c_{{\alpha}k_{v}}^{\dagger}c_{{\alpha}k_{v}}$
is the $v$th electronic reservoir. $c_{{\alpha}k_{v}}^{\dagger}(c_{{\alpha}k_{v}})$ is the creation (annihilation) operator of the electron with energy $\varepsilon_{\alpha{k_{v}}}$, chemical potential $\mu_{v\alpha}$, momentum $k_{v}$ and spin $\alpha$. The tunneling between the reservoirs and $N$-site central system is described by $V_T=\sum\limits_{{\alpha}k_L}\gamma_{L\alpha}c_{{\alpha}k_L}^{\dagger}c_{{\alpha}1}+\sum\limits_{{\alpha}k_R}\gamma_{R\alpha}c_{{\alpha}k_R}^{\dagger}c_{{\alpha}N}+h.c.$.
In the central system, besides the electron hopping terms, the localized spins $\bm{S}_i$ and $\bm{S}_j$ are embedded in the electron network through $s$-$d$ interaction. The local-spin-electron hybridized Hamiltonian is then $H_S=H_e+H_{sd}$.

The central electron hopping system may contain the Rashba spin-orbit coupling (RSOC) as $H_e=H_0+H_{SOC}$.
\begin{eqnarray}
\begin{split}
H_0=&\sum\limits_{{\alpha}i}[{\varepsilon}c_{{\alpha}i}^{\dagger}c_{{\alpha}i}+(tc_{{\alpha}i}^{\dagger}c_{{\alpha}j}+h.c.)], \\
=&\bm{c}^\dagger[(\varepsilon\bm{I}+t\bm{K})\otimes\sigma^0]\bm{c},\label{equ:H0}
\end{split}
\end{eqnarray}
where $\bm{c}^\dagger=(c^{\dag}_{{\uparrow}1},c^{\dag}_{{\downarrow}1}, ... , c^{\dag}_{{\uparrow}N}, c^{\dag}_{{\downarrow}N})$, with $c_{{\alpha}i}^{\dagger}(c_{{\alpha}i})$ being the creation (annihilation) operator of electron at $i$th site with spin $\alpha$.
For simplicity, we assume the on-site potential $\varepsilon$ and the nearest neighbor hopping integral $t$ are homogeneous. $\bm{I}$ is $N\times N$ identity matrix and $\bm K$ is the symmetric electron network with $K_{ij}=K_{ji}=1$ if there exists a hopping connection between sites $i, j$, otherwise $0$. Without loss of generality, we assume $H_{SOC}$ describes the RSOC Hamiltonian~\cite{TS2016,Bychkov1984,Christopher2016,VG2005,RM2010,YO2010},  as
\begin{eqnarray}{\label{equ:HRX}}
\begin{split}
H_{SOC}=&\sum\limits_{i}[i{\delta}c^{\dag}_{{\uparrow}i}c_{{\downarrow}j}+i{\delta}c^{\dag}_{{\downarrow}i}c_{{\uparrow}j}+h.c.],\\
=&\bm{c}^\dagger[\delta\bm{A}\otimes\sigma^x]\bm{c},
\end{split}
\end{eqnarray}
where $\bm A$ is the antisymmetric hermitian matrix that if $A_{ij}=i$ then $A_{ji}=-i$ when RSOC exists between sites $i, j$, and $\delta$ is the RSOC strength.

The interaction between the localized spin $\bm{S}_i$ and the conduction electron is described by the $s$-$d$ interaction:
\begin{eqnarray}
\begin{split}
H_{sd}=&-\frac{J}{2}\sum\limits_{{\alpha\beta}i}(c_{{\alpha}i}^{\dagger}\bm{\sigma}_{\alpha\beta}c_{{\beta}i})\cdot\bm{S}_{i},
\end{split}
\end{eqnarray}
where $J$ is the coupling strength of the $s$-$d$ interaction and $\bm{\sigma}$ is the vector of Pauli matrices.

Our goal is to obtain the indirect exchange interaction among local spins, so that we need trace out the electronic degree of freedom. As such, the spin-spin exchange interaction Hamiltonian can be written as
\begin{equation}\label{equ:Esd1}
H_{spin}=\langle H_{sd}\rangle=-\frac{J}{2}\sum\limits_{i\alpha\beta}\langle c_{{\alpha}i}^{\dagger}c_{{\beta}i}\rangle\bm{\sigma}_{\alpha\beta}\cdot\bm{S}_{i},
\end{equation}
by averaging out the electron degree of freedom, denoted as $\langle...\rangle$.
By using the Keldysh nonequilibrium Green's functions (GF)\cite{SW1996}, the Eq. (\ref{equ:Esd1}) can be expressed as
\begin{equation}\label{equ:Esd2}
H_{spin}=(-i)(-\frac{J}{2})\sum\limits_{i\alpha\beta}\bm{\sigma}_{\alpha\beta}\cdot\bm{S}_{i}\int^{\infty}_{-\infty}\frac{d\omega}{2\pi}
G_{{\beta}i,{\alpha}i}^{<}(\omega),
\end{equation}
where $G_{{\beta}i,{\alpha}i}^{<}(\omega)$ is the Fourier transform of the lesser GF
\begin{eqnarray}\label{equ:G<}
G_{{\beta}j,{\alpha}i}^{<}(t)=i\langle c_{{\alpha}i}^{\dagger}(0)c_{{\beta}j}(t)\rangle.
\end{eqnarray}
Since we treat the effective spin exchange Hamiltonian $H_{spin}$ perturbatively up to the second order of $J$, we only need to treat the nonequilibrtium GF under the first-order perturbation,
\begin{equation}\label{equ:Gw}
\begin{aligned}
G^{<}_{{\beta}i,{\alpha}i}(\omega)=&G_{{\beta}i,{\alpha}i}^{<(0)}(\omega)\\
+&(-\frac{J}{2})\sum\limits_{j\alpha'\beta'}\bm{\sigma}_{\alpha'\beta'}\cdot\bm{S}_{j}
[G_{{\beta}i,{\alpha'}j}^{(0)}(\omega)G_{{\beta'}j,{\alpha}i}^{(0)}(\omega)]^{<},
\end{aligned}
\end{equation}
where $G_{{\beta}i,{\alpha}i}^{<(0)}(\omega)$ is the Fourier transform of $G_{{\beta}j,{\alpha}i}^{<(0)}(t)=i\langle c_{{\alpha}i}^{\dagger}(0)c_{{\beta}j}(t)\rangle_0$. The superscript $(0)$ denotes the nonequilibrium GF is for the unperturbed electronic system without coupling to localized spins, but only including the pure electronic Hamiltonian $H_e+V_T+\sum_{v=L,R}H_v$. In the same manner, the $\langle ... \rangle_0$ means the average on the pure electronic system without local-spin-electron hybridization.

The first term in Eq.~(\ref{equ:Gw}) contributes to the local exchange interaction of the electron spin and the local spin, where after the perturbative treatment the electron spin behaves as the effective local magnetic field on each local spin. Meanwhile, the second term will contribute to the indirect exchange interaction among different local spins, resulting from exchanging the itinerant electrons. In the present study, we focus on the exchange interactions among different local spins that are of prime interest, so we substitute the second term of Eq.~(\ref{equ:Gw}) into Eq.~(\ref{equ:Esd2}). We finally obtain the exchange interaction Hamiltonian between $\bm{S}_i$ and $\bm{S}_j$ (see Appendix~\ref{appendixA})
\begin{equation}\label{equ:Esd3}
\begin{aligned}
H_{spin}^{ij}=&-i\frac{J^2}{4}\int^{\infty}_{-\infty}\frac{d\omega}{2\pi}\{\mathrm{Tr}\{
\bm{S}_{i}\cdot[(\bm{\sigma}\bm{G^{(0)}_{ij}})(\bm{\sigma}\bm{G^{(0)}_{ji}})]^{<}\cdot\bm{S}_{j}\\
&+\bm{S}_{j}\cdot[(\bm{\sigma}\bm{G^{(0)}_{ji}})(\bm{\sigma}\bm{G^{(0)}_{ij}})]^{<}\cdot\bm{S}_{i}\}.
\end{aligned}
\end{equation}
Here, $\bm{G^{(0)}_{ij}}$ is the spin space matrix GF for the propagator of an electron from $j$th site to $i$th site. $\mathrm{Tr}$ is the spin space trace and $(\bm{\sigma}\bm{G^{(0)}_{ij}})(\bm{\sigma}\bm{G^{(0)}_{ji}})$ is the dyad defined as $\bm{u}\bm{v}=\sum_{a,b=x,y,z}u_av_b\hat{\bm{e}}_a\hat{\bm{e}}_b$. %and $\hat{\bm{e}}_x=\hat{\bm{i}},\hat{\bm{e}}_y=\hat{\bm{j}},\hat{\bm{e}}_z=\hat{\bm{k}}$.}

After a careful derivation and reorganization, we find the spin Hamiltonian can be expressed as three parts:
\begin{equation}\label{equ:Esd3}
\begin{aligned}
H_{spin}^{ij}=&\sum\limits_{a=x,y,z}J^{a}_{ij}S^a_iS^a_j+\bm{D}_{ij}\cdot(\bm{S}_{i}\times\bm{S}_{j})+d^x_{ij}(S^y_iS^z_j+\\
&S^z_iS^y_j)+d^y_{ij}(S^x_iS^z_j+S^z_iS^x_j)+d^z_{ij}(S^x_iS^y_j+S^y_iS^z_j),
\end{aligned}
\end{equation}
where the first sum of Eq. (\ref{equ:Esd3}) describes the anisotropic Heisenberg interactions, the second term is the antisymmetric helical DM interaction and the others ones are the symmetric helical KSEA interactions. The detailed expressions of all coefficients are given by
\begin{equation}\label{equ:JD}
\begin{aligned}
J^{a}_{ij}=&-i\frac{J^2}{4}\int^{\infty}_{-\infty}\frac{d\omega}{2\pi}\mathrm{Tr}\\
&\{[(\sigma^{a}\bm{G^{(0)}_{ij}})(\sigma^{a}\bm{G^{(0)}_{ji}})]^{<}
+[(\sigma^{a}\bm{G^{(0)}_{ji}})(\sigma^{a}\bm{G^{(0)}_{ij}})]^{<}\},\\
D^{a}_{ij}=&-i\frac{J^2}{8}\sum\limits_{b,c}^{x,y,z}\varepsilon_{abc}\int^{\infty}_{-\infty}\frac{d\omega}{2\pi}\mathrm{Tr}\\
&\{[(\sigma^{b}\bm{G^{(0)}_{ij}})(\sigma^{c}\bm{G^{(0)}_{ji}})]^{<}
+[(\sigma^{c}\bm{G^{(0)}_{ji}})(\sigma^{b}\bm{G^{(0)}_{ij}})]^{<}\},\\
d^{a}_{ij}=&-i\frac{J^2}{8}\sum\limits_{b,c}^{x,y,z}\tilde{{\varepsilon}}_{abc}\int^{\infty}_{-\infty}\frac{d\omega}{2\pi}\mathrm{Tr}\\
&\{[(\sigma^{b}\bm{G^{(0)}_{ij}})(\sigma^{c}\bm{G^{(0)}_{ji}})]^{<}
+[(\sigma^{c}\bm{G^{(0)}_{ji}})(\sigma^{b}\bm{G^{(0)}_{ij}})]^{<}\}, \\
\end{aligned}
\end{equation}
where $\varepsilon_{abc}$ is Levi-Civita symbol and $\tilde{{\varepsilon}}_{abc}=|{\varepsilon}_{abc}|$ with $a, b, c=x, y, z$.
Clearly, besides the synthesized anisotropic Heisenberg interactions, the synthetic helical indirect exchange interactions possess two parts: antisymmetric DM interaction and symmetric KSEA interaction, which are formulated in terms of Keldysh nonequilibrium GFs. These expressions are our main results, which are independent on the choose of the central electronic Hamiltonian.

\section{Results and Discussions}

As we will see in the following, the presence of merely SOCs or spin polarized currents is able to synthesize and control the antisymmetric DM exchange interactions, along with the same direction as the spin splitting. However, the appearance of symmetric KSEA interactions requires both the SOCs and spin polarized currents, with different splitting directions.

To demonstrate these, let us consider three cases. For the first case, the system has RSOC but no spin polarized current.
The nonequilibrium GF $\bm{G^{(0)}_{ij}}$ can be obtained from the unperturbed nonequilibrium GF of the pure electronic system $H_0+\sum_{v=L,R}H_{v}+V_T$ while including the RSOC of $\sigma^x$ term into the Dyson equation, which can be expressed as $\bm{G^{(0)}_{ij}}=\mathcal{G}^{0}_{ij}\sigma^{0}+\mathcal{G}^{x}_{ij}\sigma^{x}$ (see Appendix~\ref{appendixB}). By substituting this GFs into Eq.~(\ref{equ:JD}), one obtains
\begin{equation}\label{equ:Esd6}
\begin{aligned}
D^{x}_{ij}=&\frac{J^2}{2}\int^{\infty}_{-\infty}\frac{d\omega}{2\pi}\mathrm{Tr}\{[\mathcal{G}^{0}_{ij}\mathcal{G}^{x}_{ji}-\mathcal{G}^{x}_{ij}\mathcal{G}^{0}_{ji}]^{<}\},\\
D^{y}_{ij}=&D^{z}_{ij}=d^{x}_{ij}=d^{y}_{ij}=d^{z}_{ij}=0.\\
\end{aligned}
\end{equation}
We find that the helical indirect DM interaction $D^{x}_{ij}$ exists in this case, due to the RSOC in the $x$ direction that contains the $\sigma^x$ term. If we choose the RSOC of $\sigma^y$ term in the $y$ direction, we will obtain nonzero $D^y$ but other terms zero.

For the second case, the system has spin polarized current but no RSOC. The spin polarization is in $z$ direction, which can result from the Zeeman splitting of the on-site  potential $\varepsilon_{\uparrow(\downarrow)}$, the system-reservoir coupling polarization $V_{T,\uparrow(\downarrow)}$, or the spin-polarized Fermi distributions in the electronic reservoirs. Thus, the nonequilibrium GF $\bm{G^{(0)}_{ij}}$ can be written as $\bm{G^{(0)}_{ij}}=\tilde{\mathcal{G}}^{0}_{ij}\sigma^{0}+\tilde{\mathcal{G}}^{z}_{ij}\sigma^{z}$ (see Appendix~\ref{appendixB}), so that one gets
\begin{equation}\label{equ:Esd6}
\begin{aligned}
D^{z}_{ij}=&\frac{J^2}{2}\int^{\infty}_{-\infty}\frac{d\omega}{2\pi}\mathrm{Tr}\{[\tilde{\mathcal{G}}^{0}_{ij}\tilde{\mathcal{G}}^{z}_{ji}-\tilde{\mathcal{G}}^{z}_{ij}\tilde{\mathcal{G}}^{0}_{ji}]^{<}\},\\
D^{x}_{ij}=&D^{y}_{ij}=d^{x}_{ij}=d^{y}_{ij}=d^{z}_{ij}=0.\\
\end{aligned}
\end{equation}
Therefore, as expected nonzero helical DM vector $D^{z}_{ij}$ appears in the $z$ direction that is different from the first case, and other terms are all zero.

For the third general case, the system has not only RSOC but also spin polarization with different splitting directions. Assuming RSOC contributes the $\sigma^x$ term and spin polarized current contributes the $\sigma^z$ term, they two together will produce the cross term $\sigma^y$ in the nonequilibrium GFs, so that the GF is generally expressed as $\bm{G^{(0)}_{ij}}=\mathcal{G}^{0}_{ij}\sigma^{0}+\sum_{a=x,y,z}\mathcal{G}^{a}_{ij}\sigma^{a}$. Therefore, all the indirect helical exchange interactions appearing in Eq. (\ref{equ:JD}), especially the symmetric KSEA interactions $d^a_{ij}$, will all emerge when both RSOC and spin polarized current are present with different polarization directions. In the following, we will analyze in detail an example of two-sites system, where both the analytic and numerical results will be examined to demonstrate these general discussions.

\subsection{Two-Sites System as an Example}\label{sec:result}
\begin{figure}[!htp]
\includegraphics[width=0.48\textwidth]{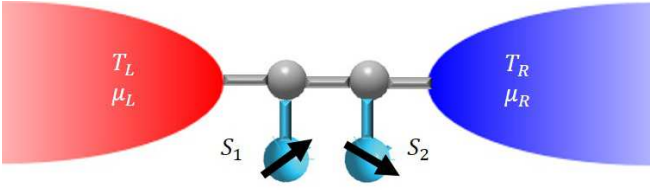}
\vspace{-0.6cm}
\caption{Example of two localized spins embedded in a two-site electron chain with the Rashba spin-orbit coupling. This chain is between the two thermalizde electronic reservoirs.}\label{fig:2}
\end{figure}

Now, without loss of generality, we take two sites as an intuitive example. Under the wide-band limit, after a standard calculation (see Appendix~\ref{appendixA}) we get the analytic expression of indirect helical exchange coefficients
\begin{eqnarray}\label{equ:JHX}
\begin{aligned}
D^{x}_{ij}=&2t{\delta}\int^\infty_{-\infty}d{\omega}(\epsilon-\omega)(f_{L{\downarrow}}+f_{L{\uparrow}}+f_{R{\downarrow}}+f_{R{\uparrow}})Q,\\
D^{y}_{ij}=&-\frac{t{\delta}{\gamma}}{2}\int^\infty_{-\infty}d{\omega}(f_{L{\downarrow}}-f_{L{\uparrow}}+f_{R{\downarrow}}-f_{R{\uparrow}})Q,\\
D^{z}_{ij}=&\frac{t^2{\gamma}}{2}\int^\infty_{-\infty}d{\omega}(f_{L{\downarrow}}-f_{L{\uparrow}}-f_{R{\downarrow}}+f_{R{\uparrow}})Q,\\
d^{x}_{ij}=&0,\\
d^{y}_{ij}=&\frac{t{\delta}{\gamma}}{2}\int^\infty_{-\infty}d{\omega}(f_{L{\downarrow}}-f_{L{\uparrow}}-f_{R{\downarrow}}+f_{R{\uparrow}})Q,\\
d^{z}_{ij}=&-\frac{{\delta}^2{\gamma}}{2}\int^\infty_{-\infty}d{\omega}(f_{L{\downarrow}}-f_{L{\uparrow}}+f_{R{\downarrow}}-f_{R{\uparrow}})Q,\\
\end{aligned}
\end{eqnarray}
with $Q=32J^2\gamma(4t^2+\gamma^2+4(\delta+\epsilon-\omega)(\delta-\epsilon+\omega))/$
$\pi(4t^2+(\gamma+2i(\delta+\epsilon-\omega))(\gamma-2i(\delta-\epsilon+\omega)))^2(4t^2+(\gamma-2i(\delta+\epsilon-\omega))
(\gamma+2i(\delta-\epsilon+\omega)))^2$, $f_{R(L)\alpha}=\frac{1}{e^{(\omega-\mu_{R(L)\alpha})/T_{R(L)}}+1}$ are the Fermi-Dirac distribution functions of the thermalized electronic reservoirs. $\mu_{R(L)\alpha}$ and $T_{R(L)}$ are the chemical potentials and the temperatures of the thermal reservoirs, respectively.

\subsection{No spin polarization}
\begin{figure}[!htp]
\centering
\includegraphics[width=0.48\textwidth]{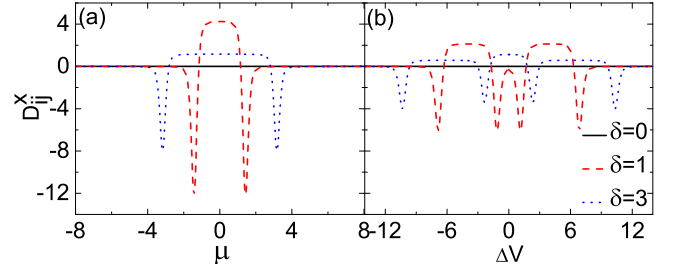}
\vspace{-1.0cm}
\caption{(colour online) DM interaction $D^{x}_{ij}$ as functions of (a) chemical potential $\mu=\mu_L=\mu_R$ and (b) bias voltage ${\Delta}V=\mu_{L}-\mu_{R}$.
The plots in (b) is $\mu_{L}=\frac{{\Delta}V}{2}+2, \mu_{R}=-\frac{{\Delta}V}{2}+2$.
Here, $\varepsilon=0meV,t=1meV,J=5meV,T_L=T_R=1K, \gamma_L=\gamma_R=\frac{1}{12}meV$, while the colors refer to different RSOC parameter $\delta$.}\label{fig:3}
\end{figure}
Firstly, we consider the system has RSOC but no spin polarization, i.e. $\mu_{L\uparrow}=\mu_{L\downarrow}=\mu_L, \mu_{R\uparrow}=\mu_{R\downarrow}=\mu_R$. From Eq. (\ref{equ:JHX}), we find that $D^{y}_{ij}=D^{z}_{ij}=d^{x}_{ij}=d^{y}_{ij}=d^{z}_{ij}=0$.
Only the term
\begin{equation}%\label{equ:JD}
\begin{aligned}
D^x_{ij}(\bm{S}_{i}\times\bm{S}_{j})_x.
\end{aligned}
\end{equation}
survives, different from that of the system without RSOC \cite{Ren2014}.

In Fig. \ref{fig:3}(a), we give the results of $D^{x}_{ij}$ as functions of chemical potential $\mu=\mu_L=\mu_R$ for different RSOC parameter $\delta$, which corresponds to the equilibrium case. We can see that $D^{x}_{ij}$ can be tuned from positive (negative) to negative (positive) as $\mu$ changes.
In Fig. \ref{fig:3}(b), we give the results of $D^{x}_{ij}$ as functions of bias voltage ${\Delta}V=\mu_{L}-\mu_{R}$ for different RSOC parameter $\delta$. There are electric currents in this nonequilibrium scenario. We can see that both the sign and magnitude of $D^{x}_{ij}$ can be controlled by nonequilibrium electric currents.

\begin{figure*}[!htp]
\includegraphics[width=0.35\textwidth]{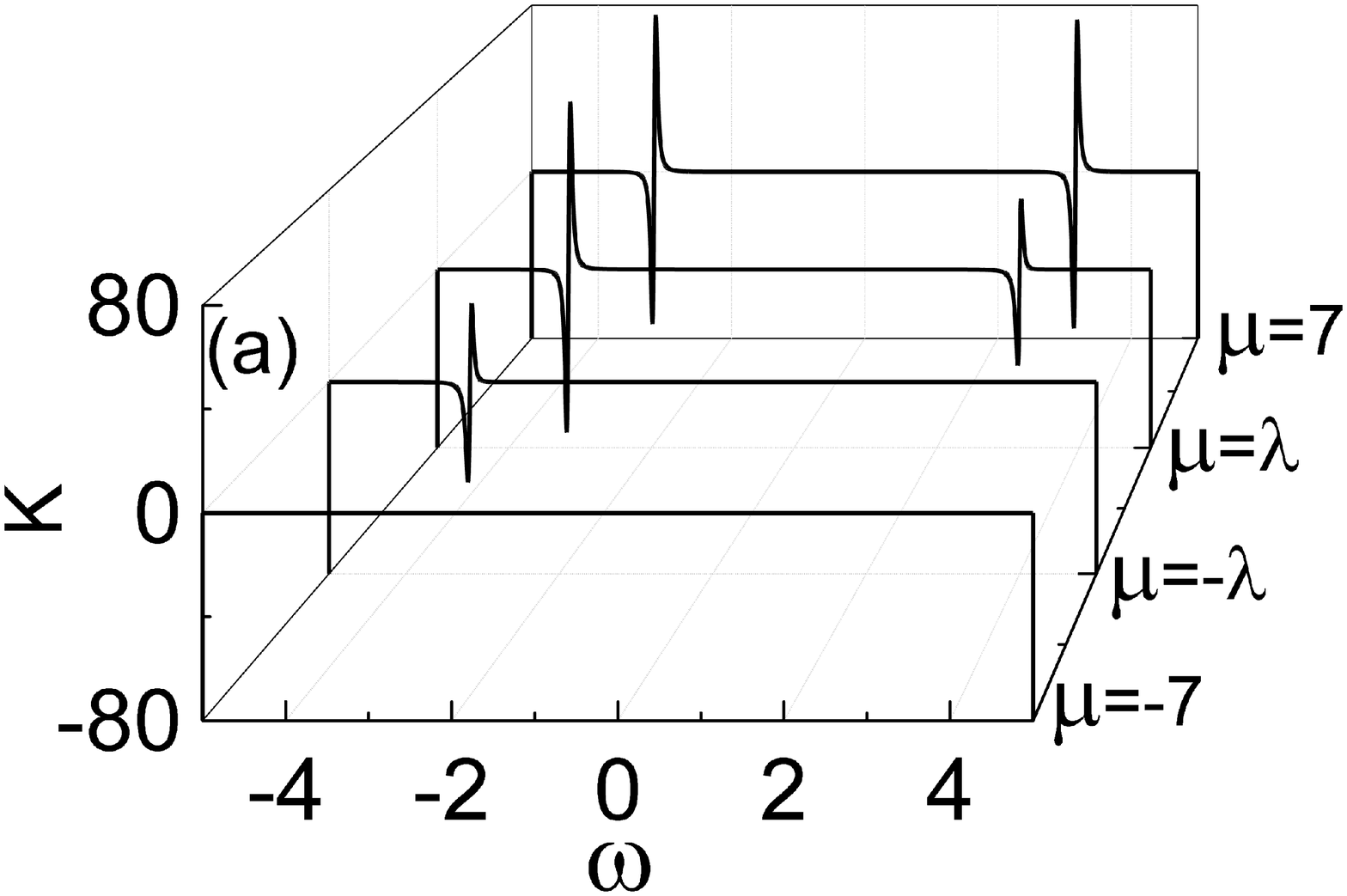}\hspace{-0.3cm}
\includegraphics[width=0.33\textwidth]{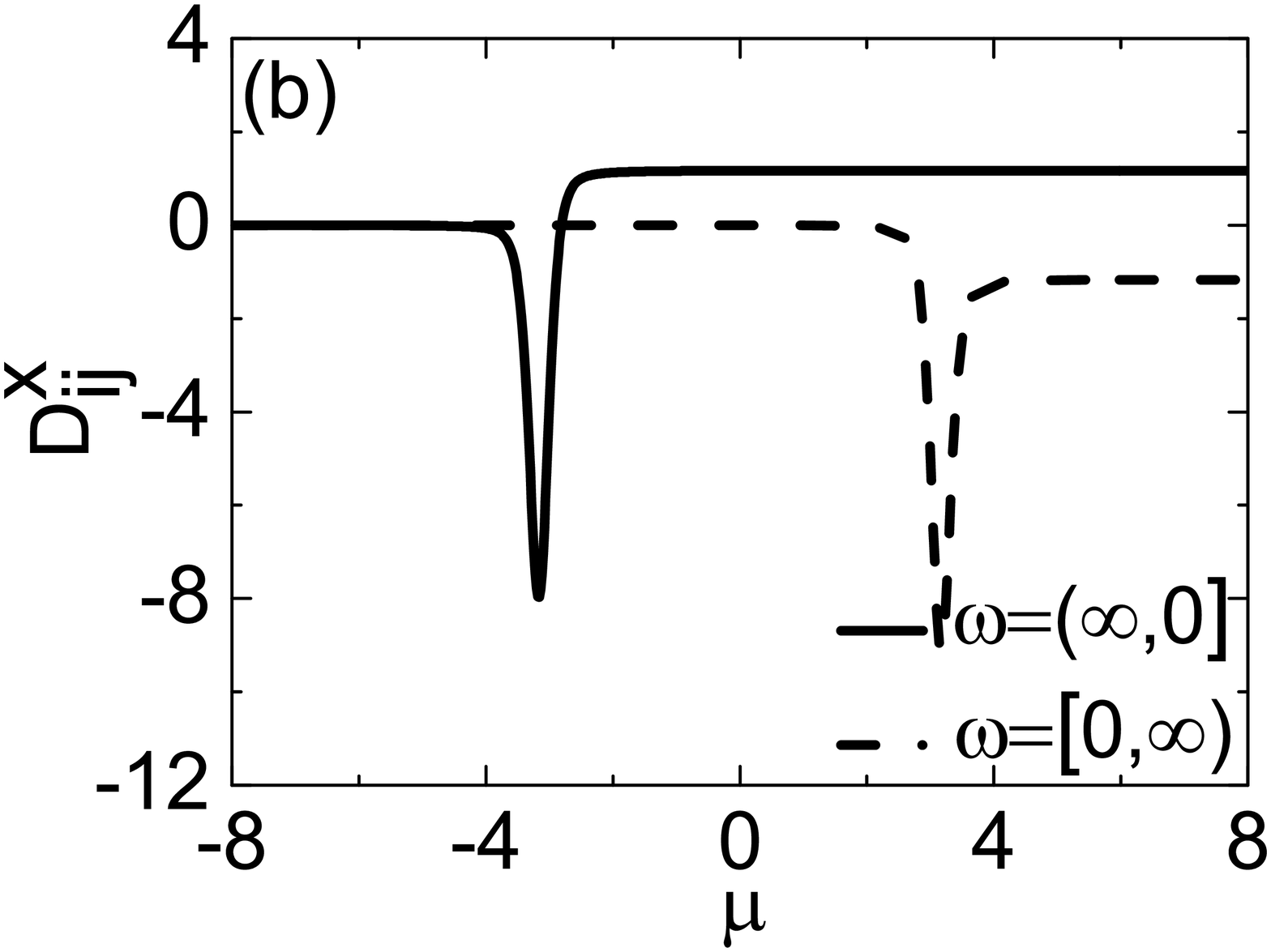}\hspace{-0.5cm}
\includegraphics[width=0.25\textwidth]{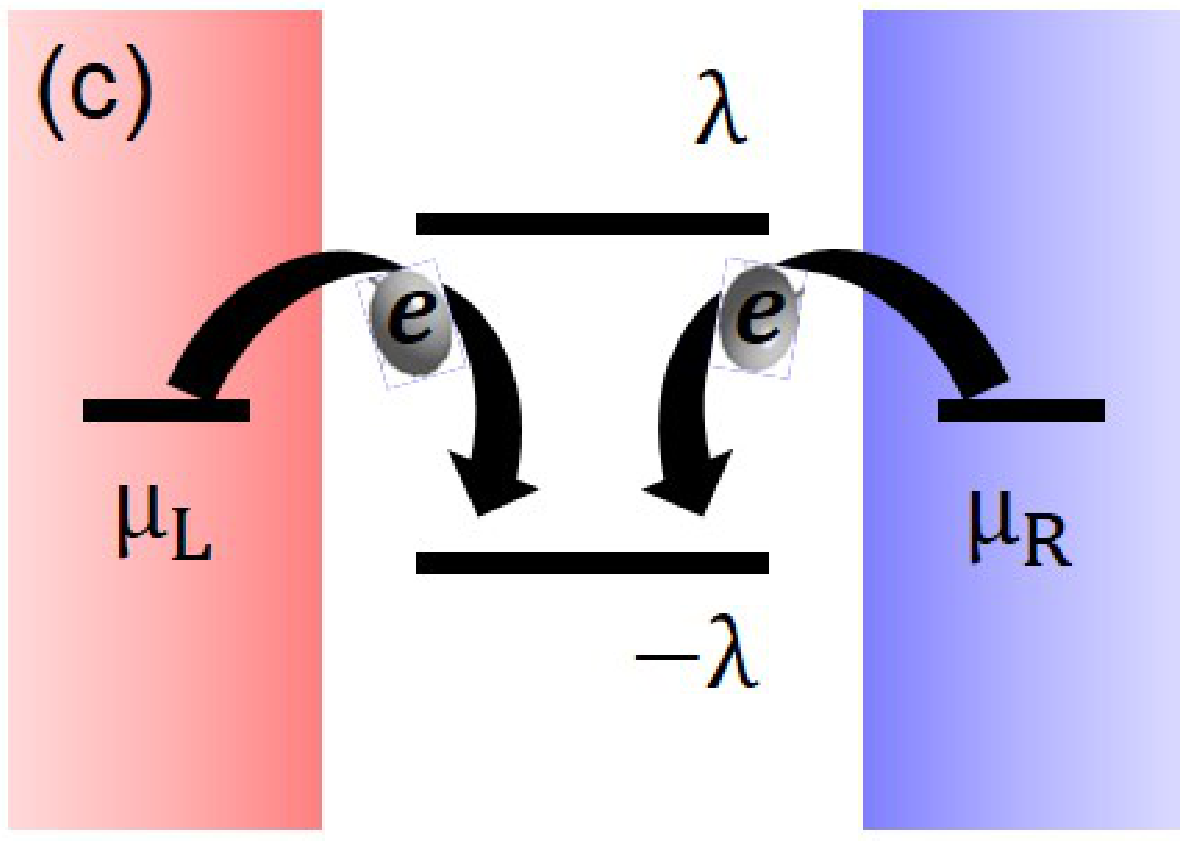}
\caption{The plots in (a) is the integrand function $K$ as functions of $\omega$ for $\mu=-7meV$, $\mu=-{\lambda}meV$, $\mu={\lambda}meV$ and $\mu=7meV$, respectively.
The plots in (b) is $D^{x}_{ij}$ as functions of $\mu$ for the bounds of integrand $\omega\in(-\infty,0]$ (solid line) and $\omega\in[0,\infty)$ (dash line), respectively.
Here, $\varepsilon=0meV, t=1meV, \delta=3meV, J=5meV, T_L=T_R=1K, \gamma_L=\gamma_R=\frac{1}{12}meV$.
The plots in (c) is schematic of how energy levels are occupied by electrons of the thermalized electronic reservoirs.}\label{fig:4}
\end{figure*}

In order to understand behaviors of $D^{x}_{ij}$, we study the integrand function of Eq. (\ref{equ:JHX}). In Fig. \ref{fig:4}(a), we plot the integrand function $K=(\epsilon-\omega)(f_{L{\downarrow}}+f_{L{\uparrow}}+f_{R{\downarrow}}+f_{R{\uparrow}})Q$ as a function of $\omega$. In this case, the energy levels of conduction electron $H_0$ are $\pm\sqrt{t^2+\delta^2}=\pm\lambda$.
When $\mu\ll-\lambda$, $K=0$. This is because that two energy levels have no electrons.
When $\mu$ closes $-\lambda$, we can see that $K$ has obvious nonzero value at $\omega=-\lambda$. This is because that the energy level $-\lambda$ start to be occupied by electrons.
When $\mu$ closes $\lambda$, the energy level $-\lambda$ is all occupied and the electrons start jump into the energy level $\lambda$.
When $\mu\gg\lambda$, the energy level $\lambda$ is fully occupied.
For further study the contribution of two energy levels to $D^{x}_{ij}$, we give the results of $D^{x}_{ij}$ as functions of $\mu$ for different range of integration $\omega\in(-\infty,0]$ and $\omega\in[0,\infty)$ in Fig. \ref{fig:4}(b).
For smaller $\mu$, two energy levels are not occupied by the electrons and the magnetic interaction is not produced. When part of two energy levels are occupied by electrons, the interaction is always negative. However, when two energy levels are all occupied by electrons, the interaction is positive for the energy level $-\lambda$ and it is negative for the energy level $\lambda$. And the positive interaction cancels with the negative interaction. For the system has bias voltage, $\mu_{L}\neq\mu_{R}$, each energy level will be filled twice by the thermalized electronic reservoirs. Therefore, $D^{x}_{ij}$ as functions of ${\Delta}V$ have four peaks (see Fig. \ref{fig:3}(b)).
Clearly, the indirect exchange interaction requires itinerate electrons rather than frozen ones.

\subsection{Spin polarized current}
\begin{figure*}[!htp]
\centering
\includegraphics[width=0.35\textwidth]{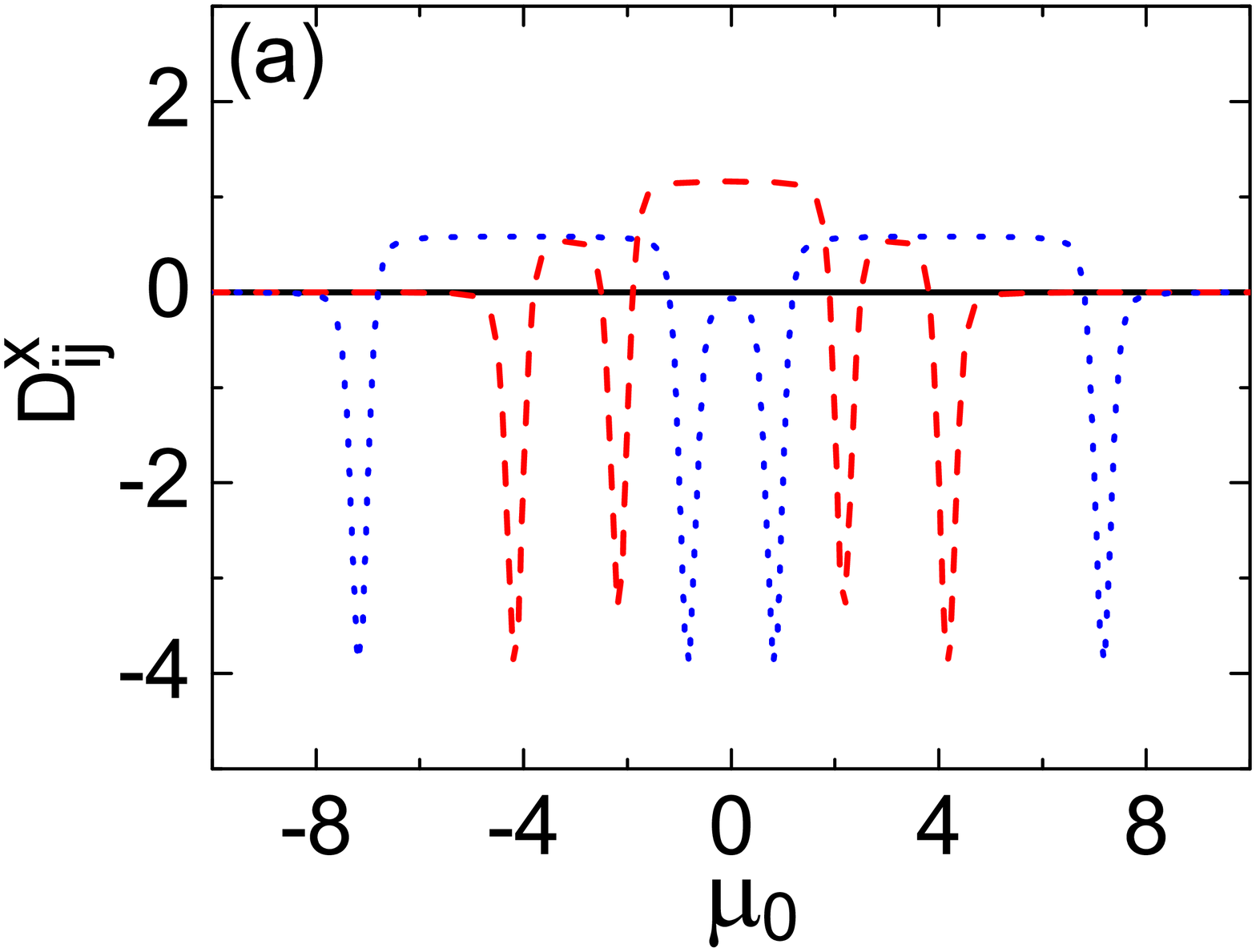}\hspace{-0.6cm}
\includegraphics[width=0.35\textwidth]{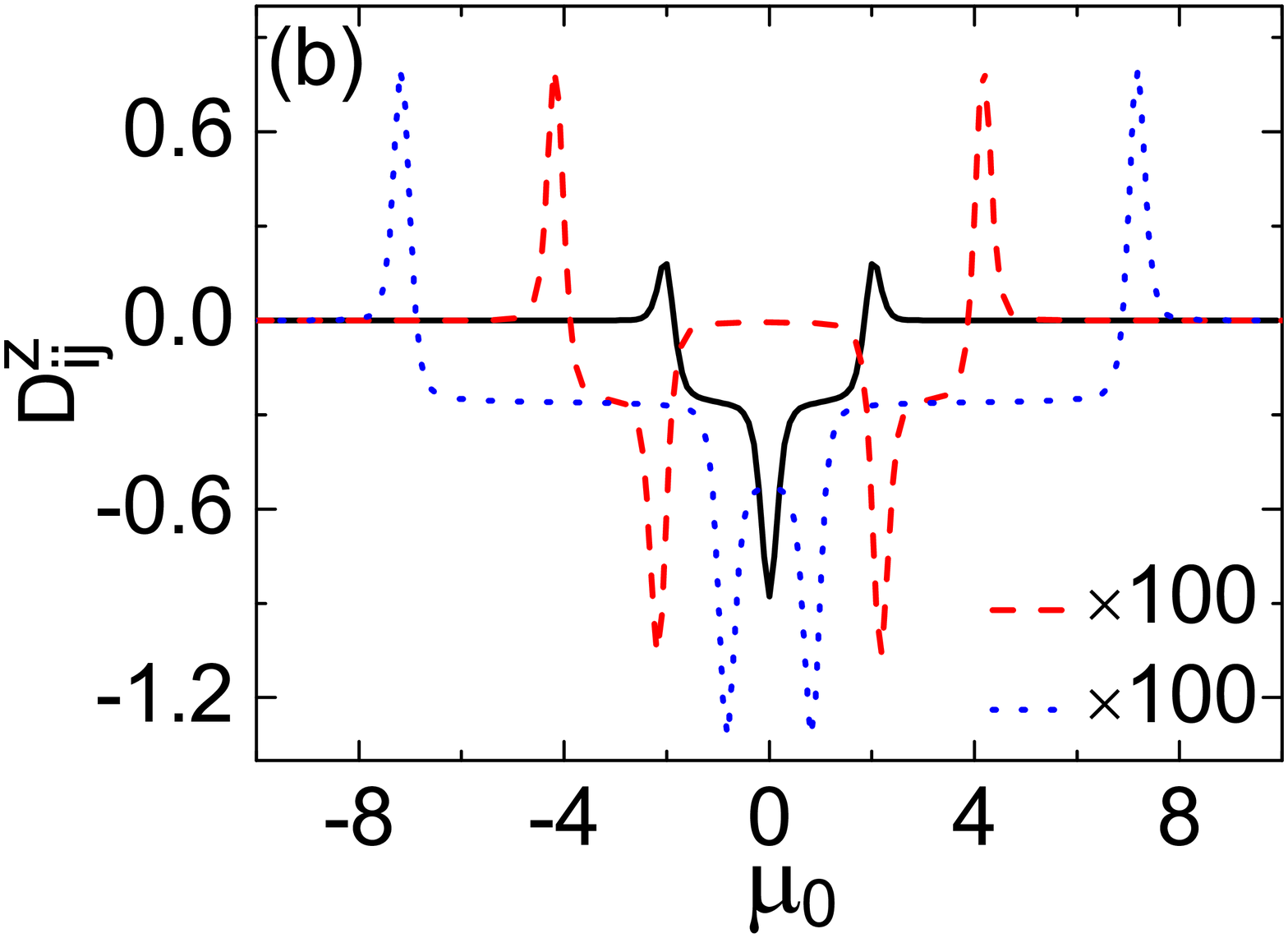}\hspace{-0.6cm}
\includegraphics[width=0.35\textwidth]{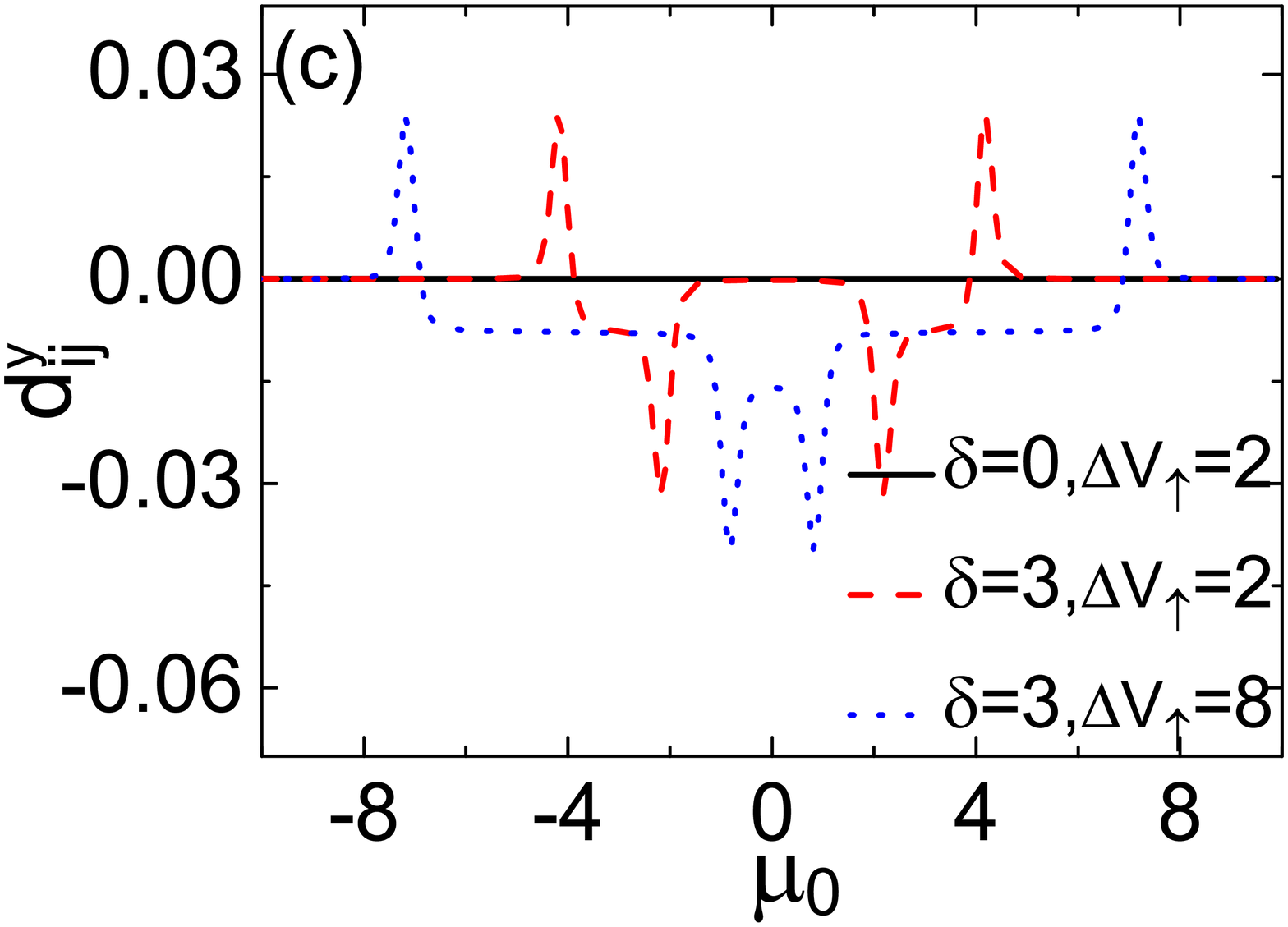}
\vspace{-0.0cm}
\caption{(colour online) Magnetic exchange interactions as functions of the average chemical potential $\mu_0$. Here, $\varepsilon=0meV,t=1meV,J=5meV,T_L=T_R=1K, \gamma_L=\gamma_R=\frac{1}{12}meV$,
${\Delta}V_\uparrow=-{\Delta}V_\downarrow$, $\mu_{L\uparrow}=\mu_{R\downarrow}=\frac{\Delta{V}_\uparrow}{2}+\mu_0$, and $\mu_{L\downarrow}=\mu_{R\uparrow}=-\frac{\Delta{V}_\uparrow}{2}+\mu_0$,
while the colors refer to different $\delta$ and $V_\uparrow$. The plots in (a) is $D^{x}_{ij}$, (b) is $D^{z}_{ij}$ and (c) is $d^{y}_{ij}$.}\label{fig:5}
\end{figure*}
In this section, we consider when the system has only spin polarized current but no electric current. So, we take the spin voltage $\mu_{L\uparrow}=\mu_{R\downarrow}$ and $\mu_{L\downarrow}=\mu_{R\uparrow}$, so that the spin voltage bias ${\Delta}V_\uparrow=-{\Delta}V_\downarrow$ are opposite since ${\Delta}V_\uparrow:=\mu_{L\uparrow}-\mu_{R\uparrow}$ and ${\Delta}V_\downarrow:=\mu_{L\downarrow}-\mu_{R\downarrow}$. From Eq.~(\ref{equ:JHX}), we know that $D^y_{ij}=d^z_{ij}=d^x_{ij}=0$.
The nonvanishing helical terms
\begin{equation}\label{equ:Esd}
D^x_{ij}(\bm{S}_{i}\times\bm{S}_{j})_x+D^z_{ij}(\bm{S}_{i}\times\bm{S}_{j})_z
+d^y_{ij}(S^x_iS^z_j+S^z_iS^x_j).
\end{equation}
may survive.
In the Fig. \ref{fig:5}, we show these coefficients of DM and KSEA interactions as functions of $\mu_0$ for different $\delta$ and ${\Delta}V_\uparrow$. Here, $\mu_0$ is the averaged chemical potential of the thermalized electronic reservoir, i.e. $\mu_0=\frac{1}{2}(\mu_{\uparrow}+\mu_{\downarrow})$.
From the Fig. \ref{fig:5}, we can see that $D^{x}_{ij}, D^{z}_{ij}$ and $d^{y}_{ij}$ are nonzero for $\delta\neq0$, however, $D^{x}_{ij}=d^{y}_{ij}=0$ and only $D^{z}_{ij}\neq0$ for $\delta=0$. We find that as long as in the presence of spin polarized current, the indirect exchange interaction appears in $z$ direction,  $D^{z}_{ij}$. And the KSEA interaction $d^{y}_{ij}$ appears when both spin polarized current and RSOC exist in the system. They can be transferred from positive (negative) to negative (positive) as the $\mu_0$ increases.

\begin{figure*}[!htp]
\centering
\includegraphics[width=0.35\textwidth]{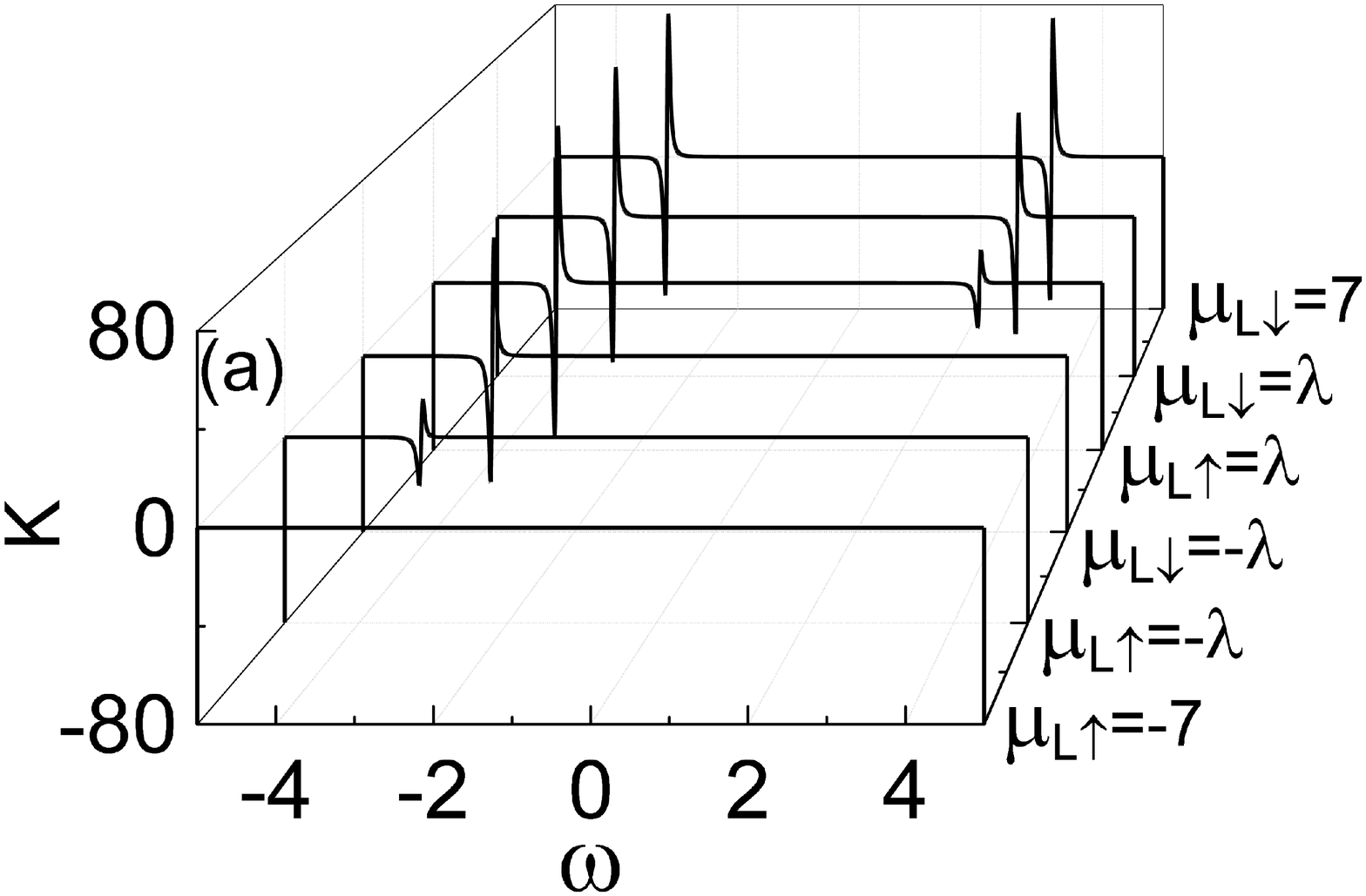}\hspace{-0.2cm}
\includegraphics[width=0.35\textwidth]{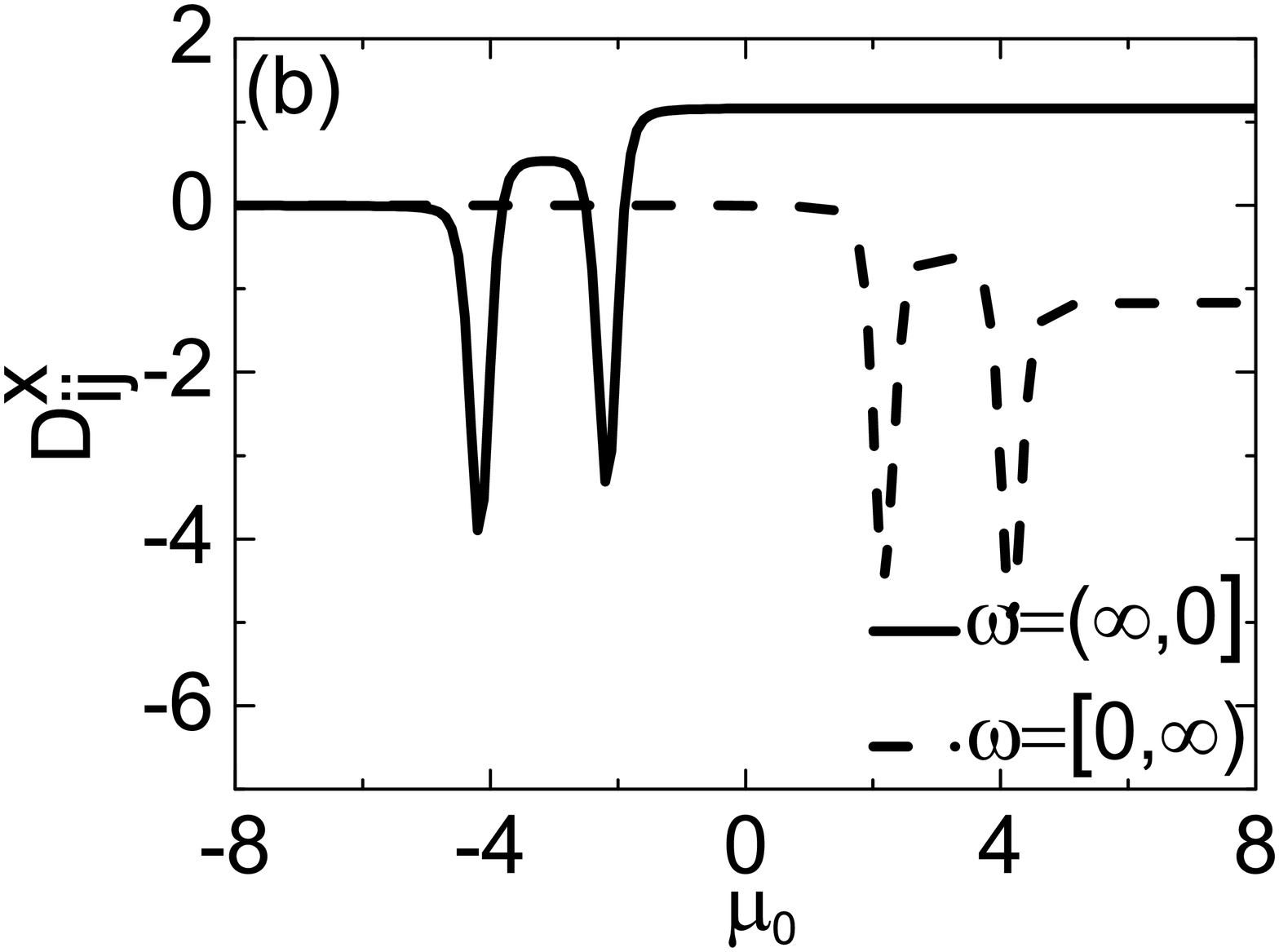}\hspace{-0.5cm}
\includegraphics[width=0.25\textwidth]{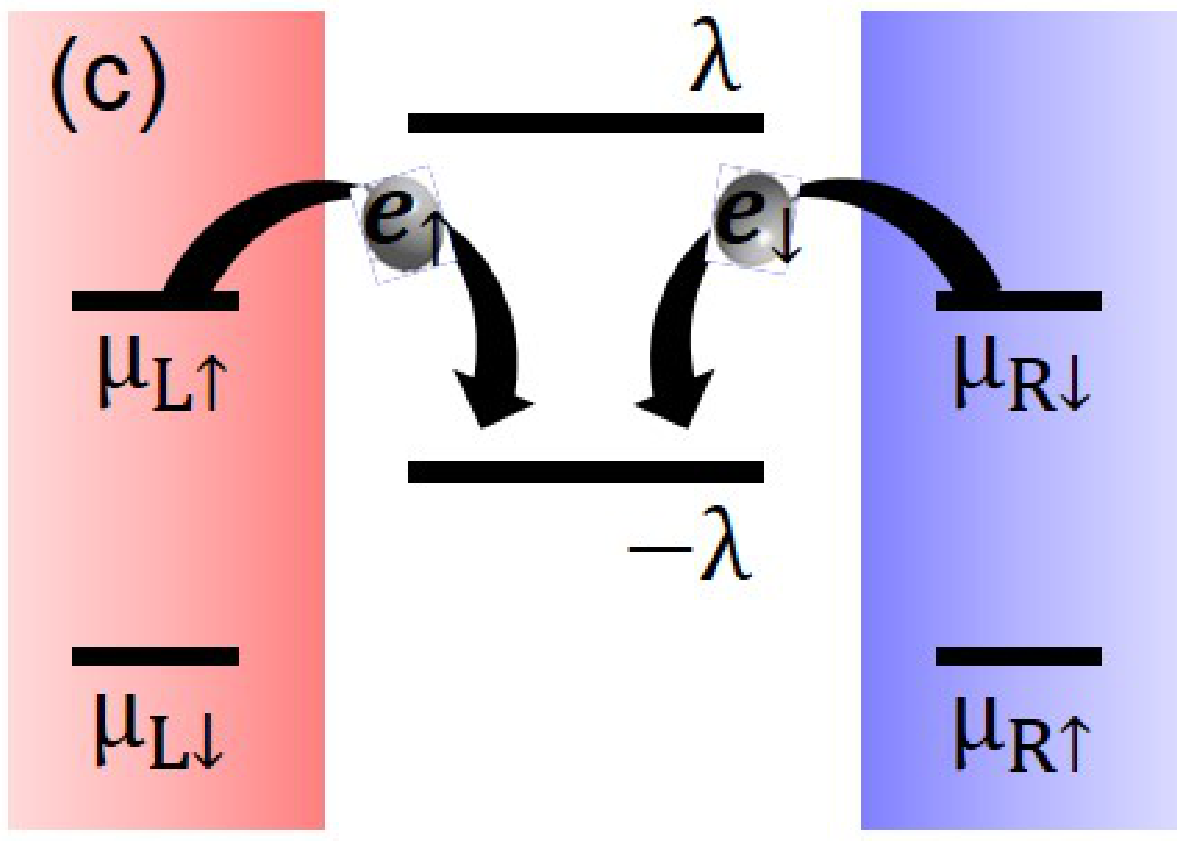}
\vspace{-0.0cm}
\caption{(colour online) The plots in (a) is the integrand function $K$ as functions of $\omega$ for $\mu_{L\uparrow}=\mu_{R\downarrow}=-7meV$, $\mu_{L\uparrow}=\mu_{R\downarrow}=-{\lambda}$, $\mu_{L\downarrow}=\mu_{R\uparrow}=-{\lambda}$, $\mu_{L\uparrow}=\mu_{R\downarrow}={\lambda}$, $\mu_{L\downarrow}=\mu_{R\uparrow}={\lambda}$ and $\mu_{L\downarrow}=\mu_{R\uparrow}=7meV$, respectively.
The plots in (b) is $D^{x}_{ij}$ as functions of $\mu$ for the bounds of integrand $\omega\in(-\infty,0]$ (solid line) and $\omega\in[0,\infty)$ (dash line), respectively.
Here, $\varepsilon=0meV, t=1meV, \delta=3meV, J=5meV, T_L=T_R=1K, \gamma_L=\gamma_R=\frac{1}{12}meV, \Delta{V}_\uparrow=2meV$.
The plots in (c) is schematic of how energy levels are occupied by electrons of the thermalized electronic reservoirs.}\label{fig:6}
\end{figure*}

Similarly, we also study the integrand function of Eq. (\ref{equ:JHX}). In Fig. \ref{fig:6}, we plot the integrand function $K$ as a function of $\omega$ and $D^{x}_{ij}$ as functions of $\mu_0$ for different range of integration $\omega\in(-\infty,0]$ and $\omega\in[0,\infty)$, respectively. Due to the system has opposite spin voltage bias, each energy level will be filled first by spin up electron of left reservoir and spin down electron of right reservoir and then by spin down electron of left reservoir and spin up electron of right reservoir. Therefore, $D^{x}_{ij}$ have four peaks (corresponding to Fig. \ref{fig:5}(a)). The behaviors of $D^{z}_{ij}$ and $d^{y}_{ij}$ can be understood by similar arguments.

\subsection{Spin polarized and electric currents}
\begin{figure}[!htp]
\centering
\includegraphics[width=0.48\textwidth]{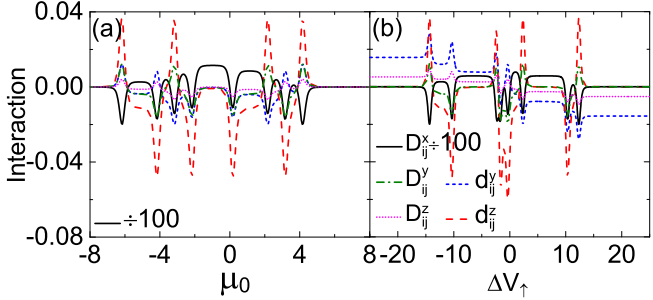}
\vspace{-1.2cm}
\caption{(colour online)  Magnetic exchange interactions as functions of (a) average chemical potential $\mu_0$ when $\Delta{V}_\uparrow=2meV$ and (b) spin voltage bias ${\Delta}V_\uparrow$ when $\mu_0=2meV$. Here, $\varepsilon=0meV,t=1meV,\delta=3meV,J=5meV,T_L=T_R=1K, \gamma_L=\gamma_R=\frac{1}{12}meV$, $\Delta{V}_\uparrow{\neq}\Delta{V}_\downarrow$ and $\mu_{L\uparrow}=\frac{\Delta{V}_\uparrow}{2}+\mu_0,
\mu_{L\downarrow}=-\frac{\Delta{V}_\uparrow}{2}+\mu_0+1, \mu_{R\uparrow}=-\frac{\Delta{V}_\uparrow}{2}+\mu_0, \mu_{R\downarrow}=\frac{\Delta{V}_\uparrow}{2}+\mu_0+2$. And the colors refer to different magnetic exchange interactions.}\label{fig:7}
\end{figure}
Finally, we consider that the system has both the electric current and the spin polarized current, where the helical DM and KSEA interactions are not zero. In the Fig. \ref{fig:7}, we show these coefficients of DM and KSEA type interactions as functions of $\mu_0$ and $\Delta V_{\uparrow}$, respectively. We can see that these interactions have eight peaks, corresponding to the eight ways for filling levels of four different electrons, i.e. the left electron with spin up or spin down, the right electron with spin up or spin down.

\section{SUMMARY}\label{sec:summary}
In summary we have studied the nonequilibrium helical indirect exchange interactions of local spins that embedded in general open electronic systems.
We have found that besides the synthesized anisotropic Heisenberg interactions, there also appear the synthetic helical indirect exchange interactions: antisymmetric DM interaction and symmetric KSEA interaction.
When the system only has spin-orbit coupling or spin polarized currents, it can synthesize and control the antisymmetric DM exchange interactions, as the same direction as spin splitting.
However, the appearance of symmetric KSEA interactions requires the system has not only the spin-orbit coupling but also spin polarized currents with different splitting directions.
Moreover, we can control the sign, magnitude, and direction of indirect DM vectors and KSEA interactions by bias voltage and spin voltage bias at nonequilibrium in open quantum devices. These results and the framework may find potential applications in  magnonics and spintronics, where the helical DM and KSEA interactions are important to control the order symmetry of spin spiral and spin texture.

\begin{appendix}
\section{GF details for the helical interactions}\label{appendixA}
In order to calculate lesser GF $G^{<}_{{\beta}i,{\alpha}i}(\omega)$ of Eq. (\ref{equ:Esd2}), we defined the nonequilibrtium GF as
\begin{eqnarray}\label{aequ:}
G_{{\beta}i,{\alpha}i}(t,t')=-i{\langle}T_ce^{-i{\int}_cd{\tau}H_{sd}({\tau})}c_{{\beta}i}(t)c_{{\alpha}i}^{\dagger}(t')\rangle,
\end{eqnarray}
since we treat the effective spin exchange Hamiltonian $H_{spin}$ perturbatively up to the second order of $J$, we only need to treat the nonequilibrtium GF under the first-order perturbation,
\begin{eqnarray}\label{aequ:}
\begin{aligned}
G_{{\beta}i,{\alpha}i}(t,t')=&G^{(0)}_{{\beta}i,{\alpha}i}(t,t')+(-\frac{J}{2})\sum\limits_{j\alpha'\beta'}\bm{\sigma}_{\alpha'\beta'}\cdot\bm{S}_{j}\\
&{\int}_cd{t_1}G^{(0)}_{{\beta}i,{\alpha'}j}(t,t_1)G^{(0)}_{{\beta'}j,{\alpha}i}(t_1,t'),
\end{aligned}
\end{eqnarray}
where $G_{{\beta}j,{\alpha}i}^{(0)}(t)=i<c_{{\alpha}i}^{\dagger}(0)c_{{\beta}j}(t)>_0$.
The $\langle ... \rangle_0$ is the average on the unperturbed electronic system without coupling to localized spins, but only including the pure electronic Hamiltonian $H_e+V_T+\sum_{v=L,R}H_v$.
By taking Fourier transform, we get
\begin{equation}\label{aequ:}
\begin{aligned}
G_{{\beta}i,{\alpha}i}(\omega)=&G_{{\beta}i,{\alpha}i}^{(0)}(\omega)+\\
&(-\frac{J}{2})\sum\limits_{j\alpha'\beta'}\bm{\sigma}_{\alpha'\beta'}\cdot\bm{S}_{j}
G_{{\beta}i,{\alpha'}j}^{(0)}(\omega)G_{{\beta'}j,{\alpha}i}^{(0)}(\omega),
\end{aligned}
\end{equation}
and $G^{<}_{{\beta}i,{\alpha}i}(\omega)$ of Eq. (\ref{equ:Esd2}) can be written as
\begin{equation}\label{aequ:Gw}
\begin{aligned}
G^{<}_{{\beta}i,{\alpha}i}(\omega)=&G_{{\beta}i,{\alpha}i}^{<(0)}(\omega)\\
+&(-\frac{J}{2})\sum\limits_{j\alpha'\beta'}\bm{\sigma}_{\alpha'\beta'}\cdot\bm{S}_{j}
[G_{{\beta}i,{\alpha'}j}^{(0)}(\omega)G_{{\beta'}j,{\alpha}i}^{(0)}(\omega)]^{<}.
\end{aligned}
\end{equation}

In the present study, we focus on the exchange interactions among different local spins that are of prime interest, so we substitute the second term of Eq. (\ref{aequ:Gw}) into Eq. (\ref{equ:Esd2}). We finally obtain the exchange interaction Hamiltonian between $\bm{S}_i$ and $\bm{S}_j$
\begin{equation}\label{aequ:Esd1}
\begin{aligned}
H_{spin}^{ij}
=&-i\frac{J^2}{4}\sum\limits_{i<j}\sum\limits_{\alpha\beta}\sum\limits_{\alpha'\beta'}\{(\bm{\sigma}_{\alpha\beta}\cdot\bm{S}_{i})(\bm{\sigma}_{\alpha'\beta'}\cdot\bm{S}_{j})\\
&\int^{\infty}_{-\infty}\frac{d\omega}{2\pi}[G_{{\beta}i,{\alpha'}j}^{(0)}(\omega)G_{{\beta'}j,{\alpha}i}^{(0)}(\omega)]^{<}\\
&+(\bm{\sigma}_{\alpha\beta}\cdot\bm{S}_{j})(\bm{\sigma}_{\alpha'\beta'}\cdot\bm{S}_{i})\\
&\int^{\infty}_{-\infty}\frac{d\omega}{2\pi}[G_{{\beta}j,{\alpha'}i}^{(0)}(\omega)G_{{\beta'}i,{\alpha}j}^{(0)}(\omega)]^{<}\},\\
=&-i\frac{J^2}{4}\sum\limits_{a,b}^{x,y,z}\sum\limits_{i<j}\sum\limits_{\alpha\beta}\sum\limits_{\alpha'\beta'}\int^{\infty}_{-\infty}\frac{d\omega}{2\pi}\\
(&\sigma^a_{\alpha\beta}\sigma^b_{\alpha'\beta'}{S}^a_{i}{S}^b_{j}[G_{{\beta}i,{\alpha'}j}^{(0)}(\omega)G_{{\beta'}j,{\alpha}i}^{(0)}(\omega)]^{<}+\\
&\sigma^a_{\alpha\beta}\sigma^b_{\alpha'\beta'}{S}^a_{j}{S}^b_{i}[G_{{\beta}j,{\alpha'}i}^{(0)}(\omega)G_{{\beta'}i,{\alpha}j}^{(0)}(\omega)]^{<}),\\
=&-i\frac{J^2}{4}\int^{\infty}_{-\infty}\frac{d\omega}{2\pi}\{\mathrm{Tr}\{
\bm{S}_{i}\cdot[(\bm{\sigma}\bm{G^{(0)}_{ij}})(\bm{\sigma}\bm{G^{(0)}_{ji}})]^{<}\cdot\bm{S}_{j}\\
&+\bm{S}_{j}\cdot[(\bm{\sigma}\bm{G^{(0)}_{ji}})(\bm{\sigma}\bm{G^{(0)}_{ij}})]^{<}\cdot\bm{S}_{i}\},\\
\end{aligned}
\end{equation}
where, $\bm{G^{(0)}_{ij}}$ is the spin space matrix GF for propagation of an electron from $jth$ site to $ith$ site and $(\bm{\sigma}\bm{G^{(0)}_{ij}})(\bm{\sigma}\bm{G^{(0)}_{ji}})$ is the dyad.

The spin Hamiltonian can be expressed as three parts
\begin{equation}\label{aequ:Esd2}
\begin{aligned}
H_{spin}^{ij}=&\sum\limits_{a=x,y,z}J^{a}_{ij}S^a_iS^a_j+\bm{D}_{ij}\cdot(\bm{S}_{i}\times\bm{S}_{j})+d^x_{ij}(S^y_iS^z_j+\\
&S^z_iS^y_j)+d^y_{ij}(S^x_iS^z_j+S^z_iS^x_j)+d^z_{ij}(S^x_iS^y_j+S^y_iS^z_j),
\end{aligned}
\end{equation}
where the first sum of Eq. (\ref{aequ:Esd2}) describes the anisotropic Heisenberg interactions, the second term is the antisymmetric DM interaction and the others ones are the symmetric KSEA interactions. The detailed expressions of all coefficients are given by
\begin{equation}\label{aequ:JD}
\begin{aligned}
J^{a}_{ij}=&-i\frac{J^2}{4}\int^{\infty}_{-\infty}\frac{d\omega}{2\pi}\mathrm{Tr}\\
&\{[(\sigma^{a}\bm{G^{(0)}_{ij}})(\sigma^{a}\bm{G^{(0)}_{ji}})]^{<}
+[(\sigma^{a}\bm{G^{(0)}_{ji}})(\sigma^{a}\bm{G^{(0)}_{ij}})]^{<}\},\\
D^{a}_{ij}=&-i\frac{J^2}{8}\sum\limits_{b,c}^{x,y,z}\varepsilon_{abc}\int^{\infty}_{-\infty}\frac{d\omega}{2\pi}\mathrm{Tr}\\
&\{[(\sigma^{b}\bm{G^{(0)}_{ij}})(\sigma^{c}\bm{G^{(0)}_{ji}})]^{<}
+[(\sigma^{c}\bm{G^{(0)}_{ji}})(\sigma^{b}\bm{G^{(0)}_{ij}})]^{<}\},\\
d^{a}_{ij}=&-i\frac{J^2}{8}\sum\limits_{b,c}^{x,y,z}\tilde{{\varepsilon}}_{abc}\int^{\infty}_{-\infty}\frac{d\omega}{2\pi}\mathrm{Tr}\\
&\{[(\sigma^{b}\bm{G^{(0)}_{ij}})(\sigma^{c}\bm{G^{(0)}_{ji}})]^{<}
+[(\sigma^{c}\bm{G^{(0)}_{ji}})(\sigma^{b}\bm{G^{(0)}_{ij}})]^{<}\}, \\
\end{aligned}
\end{equation}
where $\varepsilon_{abc}$ is Levi-Civita symbol and $\tilde{{\varepsilon}}_{abc}=|{\varepsilon}_{abc}|$ with $a, b, c=x, y, z$.

We can calculate $[\bm{G^{(0)}_{ij}}\bm{G^{(0)}_{ji}}]^{<}$ by using Langreth formula,
\begin{eqnarray}
\begin{aligned}
{\ }[\bm{G^{(0)}_{ij}}\bm{G^{(0)}_{ji}}]^{<}=&\bm{G^{<(0)}_{ij}}\bm{G^{a(0)}_{ji}}+\bm{G^{r(0)}_{ij}}\bm{G^{<(0)}_{ji}},\\
\bm{G^{<(0)}}=&\bm{G^{r(0)}}\bm{\Sigma}^<\bm{G^{a(0)}}.\\
\end{aligned}
\end{eqnarray}
Where the $\bm{G^{r(0)}}$ ($\bm{G^{a(0)}}$) is the retarded (advanced) GF, and
\begin{eqnarray}
\bm{G^{r(0)}}&=(\bm{G^{a(0)}})^\dagger=(\omega\bm{I}-\bm{H_e}-\bm{\Sigma_L}-\bm{\Sigma_R})^{-1}.
\end{eqnarray}
Here, $\bm{\Sigma_{L(R)}}$ are the self-energy of the thermalized electronic reservoirs. In general, it depends on the structure of the thermal reservoirs. In the present study, we don't consider the details of the thermal reservoirs and use the wide-band limit,
\begin{eqnarray}
\bm{\Sigma_{L(R)}}=-\frac{i}{2}\bm{\Gamma_{L(R)}}, \;\;\;
\bm{\Sigma^<}=i(f_{L\alpha}\bm{\Gamma_{L}}+f_{R\alpha}\bm{\Gamma_{R}}).
\end{eqnarray}
Where, $f_{R(L)\alpha}=\frac{1}{e^{(\omega-\mu_{R(L)\alpha})/T_{R(L)}}+1}$ are the Fermi-Dirac distribution functions of the thermal reservoirs. $\mu_{R(L)\alpha}$, $T_{R(L)}$ and $\bm{\Gamma_{L(R)}}$ are the chemical potentials, the temperatures and the coupling matrices of the thermal reservoirs, respectively.

Here, for the system has the spin polarization, the chemical potentials of the thermal reservoirs depend on the spin of electron. It is different from that of Ref. \cite{Ren2014}, where the coupling matrices $\bm{\Gamma_{L(R)}}$ depend on the spin of electron.

\section{Dyson equations leading to general results}\label{appendixB}
For the system has RSOC but no spin polarized current, the nonequilibrium GF $\bm{G^{(0)}_{ij}}$ can be obtained from the unperturbed nonequilibrium GF $\bm{g}$ of the pure electronic system $H_0+\sum_{v=L,R}H_{v}+V_T$ by the Dyson equation:
\begin{equation}\label{aequ:G0x}
\begin{aligned}
\bm{G^{(0)}}=&\bm{g}\otimes\sigma^{0}+\bm{g}\otimes\sigma^{0}H_{SOC}\bm{g}\otimes\sigma^{0}+\bm{g}\otimes\sigma^{0}H_{SOC}\\
&\bm{g}\otimes\sigma^{0}H_{SOC}\bm{g}\otimes\sigma^{0}+\cdots,\\
=&(\bm{g}+\delta^2\bm{g}\bm{A}\bm{g}\bm{A}\bm{g}+\delta^4\bm{g}\bm{A}\bm{g}\bm{A}\bm{g}\bm{A}\bm{g}\bm{A}\bm{g}+\cdots)\otimes\sigma^0\\
&+(\delta\bm{g}\bm{A}\bm{g}+\delta^3\bm{g}\bm{A}\bm{g}\bm{A}\bm{g}\bm{A}\bm{g}+\cdots)\otimes\sigma^x,\\
=&\bm{\mathcal{G}}^{0}\otimes\sigma^{0}+\bm{\mathcal{G}}^{x}\otimes\sigma^{x}.
\end{aligned}
\end{equation}
So, the spin space matrix GF for propagation of an electron from $j$th site to $i$th site is $\bm{G^{(0)}_{ij}}=\mathcal{G}^{0}_{ij}\sigma^{0}+\mathcal{G}^{x}_{ij}\sigma^{x}$. As such, the DM interaction in $x$ direction can be obtained
\begin{equation}\label{equ:Esd5}
\begin{aligned}
D^{x}_{ij}=&-i\frac{J^2}{8}\int^{\infty}_{-\infty}\frac{d\omega}{2\pi}\mathrm{Tr}\\
&\{[\sigma^{y}(\mathcal{G}^{0}_{ij}\sigma^{0}+\mathcal{G}^{x}_{ij}\sigma^{x})
\sigma^{z}(\mathcal{G}^{0}_{ji}\sigma^{0}+\mathcal{G}^{x}_{ji}\sigma^{x})]^{<}\\
&+[\sigma^{z}(\mathcal{G}^{0}_{ji}\sigma^{0}+\mathcal{G}^{x}_{ji}\sigma^{x})\sigma^{y}(\mathcal{G}^{0}_{ij}\sigma^{0}+\mathcal{G}^{x}_{ij}\sigma^{x})]^{<}\\
&-[\sigma^{z}(\mathcal{G}^{0}_{ij}\sigma^{0}+\mathcal{G}^{x}_{ij}\sigma^{x})
\sigma^{y}(\mathcal{G}^{0}_{ji}\sigma^{0}+\mathcal{G}^{x}_{ji}\sigma^{x})]^{<}\\
&-[\sigma^{y}(\mathcal{G}^{0}_{ji}\sigma^{0}+\mathcal{G}^{x}_{ji}\sigma^{x})\sigma^{z}(\mathcal{G}^{0}_{ij}\sigma^{0}+\mathcal{G}^{x}_{ij}\sigma^{x})]^{<}\},\\
=&-i\frac{J^2}{4}\int^{\infty}_{-\infty}\frac{d\omega}{2\pi}\mathrm{Tr}\\
&\{[(i\mathcal{G}^{0}_{ij}\sigma^{x}-i\mathcal{G}^{x}_{ij}\sigma^{0})
(\mathcal{G}^{0}_{ji}\sigma^{0}+\mathcal{G}^{x}_{ji}\sigma^{x})]^{<}\\
&+[(-i\mathcal{G}^{0}_{ji}\sigma^{x}+i\mathcal{G}^{x}_{ji}\sigma^{0})
(\mathcal{G}^{0}_{ij}\sigma^{0}+\mathcal{G}^{x}_{ij}\sigma^{x})]^{<}\\
&-[(-i\mathcal{G}^{0}_{ij}\sigma^{x}+i\mathcal{G}^{x}_{ij}\sigma^{0})
(\mathcal{G}^{0}_{ji}\sigma^{0}+\mathcal{G}^{x}_{ji}\sigma^{x})]^{<}\\
&-[(i\mathcal{G}^{0}_{ji}\sigma^{x}-i\mathcal{G}^{x}_{ji}\sigma^{0})
(\mathcal{G}^{0}_{ij}\sigma^{0}+\mathcal{G}^{x}_{ij}\sigma^{x})]^{<}\},\\
=&\frac{J^2}{2}\int^{\infty}_{-\infty}\frac{d\omega}{2\pi}\mathrm{Tr}\{[\mathcal{G}^{0}_{ij}\mathcal{G}^{x}_{ji}-\mathcal{G}^{x}_{ij}\mathcal{G}^{0}_{ji}]^{<}\}.
\end{aligned}
\end{equation}
Other terms of Eq. (\ref{equ:JD}) can be obtained similarly.

For the system has spin polarized current but no RSOC. The spin polarization is in $z$ direction, which can result from the Zeeman splitting of the on-site  potential $\varepsilon_{\uparrow(\downarrow)}$, the system-reservoir coupling polarization $V_{T,\uparrow(\downarrow)}$, or the spin-polarized Fermi distributions in the electronic reservoirs. Therefore, the self-energy of the thermalized electronic reservoirs will be $\bm{\Sigma^<_{L\uparrow}}\neq\bm{\Sigma^<_{L\downarrow}}$ and $\bm{\Sigma^<_{R\uparrow}}\neq\bm{\Sigma^<_{R\downarrow}}$. And the GF $\bm{G^{(0)}}$ can be written as
\begin{equation}\label{aequ:G0z}
\begin{aligned}
\bm{G^{(0)}}=\bm{\tilde{\mathcal{G}}}^{0}\otimes\sigma^{0}+\bm{\tilde{\mathcal{G}}}^{z}\otimes\sigma^{z},
\end{aligned}
\end{equation}
where, $\bm{\tilde{\mathcal{G}}}^{0}=(\bm{G^{(0)}}_{\uparrow\uparrow}+\bm{G^{(0)}}_{\downarrow\downarrow})/2 ,\bm{\tilde{\mathcal{G}}}^{z}=(\bm{G^{(0)}}_{\uparrow\uparrow}-\bm{G^{(0)}}_{\downarrow\downarrow})/2$, and
\begin{equation*}
\bm{G^{(0)}}=\left(\begin{array}{cc}
\bm{G^{(0)}}_{\uparrow\uparrow} & \bm{G^{(0)}}_{\uparrow\downarrow} \\
\bm{G^{(0)}}_{\downarrow\uparrow} & \bm{G^{(0)}}_{\downarrow\downarrow} \\
\end{array}\right).
\end{equation*}
So, the spin space matrix GF for propagation of an electron from $j$th site to $i$th site is $\bm{G^{(0)}_{ij}}=\tilde{\mathcal{G}}^{0}_{ij}\sigma^{0}+\tilde{\mathcal{G}}^{z}_{ij}\sigma^{z}$.

\end{appendix}

\acknowledgments
{The work was supported by the National Natural Science Foundation of China (Grant Nos. 11575087, 11175087). J.R. is partially supported by the National Natural Science Foun- dation of China (No. 11775159), the Natural Science Foun- dation of Shanghai (No. 18ZR1442800), and the Opening Project of Shanghai Key Laboratory of Special Artificial Mi- crostructure Materials and Technology.}


\begin{thebibliography}{99}
\bibitem{ID1958} I. Dzyaloshinskii, J. Phys. Chem. Solids. 4, 241 (1958).
\bibitem{T1960} T. Moriya, Phys. Rev. 120, 91 (1960); T. Moriya, Phys. Rev. Lett. 4, 228 (1960).
\bibitem{PWA1959} P. W. Anderson, Phys. Rev. 115, 2 (1959).
\bibitem{MB2007} M. Bode, M. Heide, K. von Bergmann, P. Ferriani, S. Heinze, G. Bihlmayer, A. Kubetzka, O. Pietzsch, S. Bl\"{u}gel, and R. Wiesendanger, Nature (London) 447, 190 (2007).
\bibitem{PF2008} P. Ferriani, K. von Bergmann, E. Y. Vedmedenko, S. Heinze, M. Bode, M. Heide, G. Bihlmayer, S. Bl\"{u}gel, and R. Wiesendanger, Phys. Rev. Lett. 101, 027201 (2008).
\bibitem{SHP2014} S. H. Phark, J. A. Fischer, M. Corbetta, D. Sander, K. Nakamura, and J. Kirschner, Nat. Commun. 5, 5183 (2014).
\bibitem{AK2003} A. Kubetzka, O. Pietzsch, M. Bode, and R. Wiesendanger, Phys. Rev. B 67, 020401 (2003).
\bibitem{MHG2008} M. Heide, G. Bihlmayer, and S. Bl\"{u}gel, Phys. Rev. B 78, 140403 (2008).
\bibitem{SM2009} S. Meckler, N. Mikuszeit, A. Pressler, E. Y. Vedmedenko, O. Pietzsch, and R. Wiesendanger, Phys. Rev. Lett. 103, 157201 (2009).
\bibitem{GC2013} G. Chen, J. Zhu, A. Quesada, J. Li, A. T. N'Diaye, Y. Huo, T. P. Ma, Y. Chen, H. Y. Kwon, C. Won, Z. Q. Qiu, A. K. Schmid, and Y. Z. Wu, Phys. Rev. Lett. 110, 177204 (2013).
\bibitem{SE2013} S. Emori, U. Bauer, S. Ahn, E. Martinez, and G. Beach, Nat. Mater. 12, 611 (2013).
\bibitem{KR2013} K. Ryu, L. Thomas, S. Yang, and S. Parkin, Nat. Nanotechnol. 8, 527 (2013).
\bibitem{AN1989} A. N. Bogdanov and D. A. Yablonskii, Sov. Phys. JETP 68, 101 (1989).
\bibitem{AN2001} A. N. Bogdanov and U. K. R\"{o}{\ss}ler, Phys. Rev. Lett. 87, 037203 (2001).
\bibitem{SMB2009} S. M\"{u}hlbauer, B. Binz, F. Jonietz, C. Pfleiderer, A. Rosch, A. Neubauer, R. Georgii, and P. B\"{o}ni, Science 323, 915 (2009).
\bibitem{XZ2010} X. Z. Yu, Y. Onose, N. Kanazawa, J. H. Park, J. H. Han, Y. Matsui, N. Nagaosa, and Y. Tokura, Nature (London) 465, 901 (2010).
\bibitem{SS2012} S. Seki, X. Z. Yu, S. Ishiwata, and Y. Tokura, Science 336, 198 (2012).
\bibitem{AF2013} A. Fert, V. Cros, and J. Sampaio, Nat. Nanotechnol. 8, 152 (2013).
\bibitem{NN2013} N. Nagaosa and Y. Tokura, Nat. Nanotechnol. 8, 899 (2013).
\bibitem{RW2016} R. Wiesendanger, Nat. Rev. Mater. 1, 16044 (2016).
\bibitem{SH2011} S. Heinze, K. von Bergmann, M. Menzel, J. Brede, A. Kubetzka, R. Wiesendanger, G. Bihlmayer, and S. Bl\"{u}gel, Nat. Phys. 7, 713 (2011).
\bibitem{NR2013} N. Romming, C. Hanneken, M. Menzel, J. E. Bickel, B. Wolter, K. von Bergmann, A. Kubetzka, and R. Wiesendanger, Science 341, 636 (2013).
\bibitem{NR2015} N. Romming, A. Kubetzka, C. Hanneken, K. von Bergmann, and R. Wiesendanger, Phys. Rev. Lett. 114, 177203 (2015).
\bibitem{GC2015} G. Chen, A. Mascaraque, A. T. N'Diaye, and A. K. Schmid, Appl. Phys. Lett. 106, 242404 (2015).
\bibitem{WJ2015} W. Jiang, P. Upadhyaya, W. Zhang, G. Yu, M. B. Jungfleisch, F. Y. Fradin, J. E. Pearson, Y. Tserkovnyak, K. L. Wang, O. Heinonen, S. G. E. te Velthuis, and A. Hoffmann, Science 349, 283 (2015).
\bibitem{OBJ2016} O. Boulle, J. Vogel, H. Yang, S. Pizzini, D. de Souza Chaves, A. Locatelli, T. O. Mentes¸, A. Sala, L. D. Buda-Prejbeanu, O. Klein, M. Belmeguenai, Y. Roussign\'{e}, A. Stashkevich, S. M. Ch\'{e}rif, L. Aballe, M. Foerster, M. Chshiev, S. Auffret, I. M. Miron, and G. Gaudin, Nat. Nanotechnol. 11, 449 (2016).
\bibitem{CML2016} C. Moreau-Luchaire, C. Moutafis, N. Reyren, J. Sampaio, C. A. F. Vaz, N. Van Horne, K. Bouzehouane, K. Garcia, C. Deranlot, P. Warnicke, P. Wohlhter, J.-M. George, M. Wigand, J. Raabe, V. Cros, and A. Fert, Nat. Nanotechnol. 11, 444 (2016).
\bibitem{SWL2016} S. Woo, K. Litzius, B. Kr\"{u}ger, M. Y. Im, L. Caretta, K. Richter, M. Mann, A. Krone, R. M. Reeve, M. Weigand, P. Agrawal, I. Lemesh, M. A. Mawass, P. Fischer, M. Kl\"{a}ui, and G. S. D. Beach, Nat. Mater. 15, 501 (2016).
\bibitem{NE2001} N.E. Bonesteel, D. Stepanenko, and D.P. DiVincenzo, Phys. Rev. Lett. 87, 207901 (2001).
\bibitem{SC2006} S. Chutia, M. Friesen, and R. Joynt, Phys. Rev. B 73, 241304 (2006).
\bibitem{RJ2008} R. Jafari, M. Kargarian, A. Langari, and M. Siahatgar, Phys. Rev. B 78, 214414 (2008).
\bibitem{ZM2013} M. Zhong, H. Xu, X. X. Liu, and P. Q. Tong, Chin. Phys. B  22,090313 (2013).
\bibitem{LW2012} W. J. Li, Z. J. Zhang, and P. Q. Tong, Eur. Phys. J. B 85, 73 (2012).
\bibitem{ss2008} S. Seki, Y. Yamasaki, M. Soda, M. Matsuura, K. Hirota, and Y. Tokura, Phys. Rev. Lett. 100 127201 (2008).
\bibitem{lq2008} Q. C. Li, S. Dong, and J. M. Liu, Phys. Rev. B 77 054442 (2008).
\bibitem{si2006} I. A. Sergienko and E. Dagotto, Phys. Rev. B 73 094434 (2006).
\bibitem{MK2008} M. K. Kwan, Z. N. Gurkan, and L. C. Kwek, Phys. Rev. A 77, 062311 (2008).
\bibitem{MK2009} M. Kargarian, R. Jafari, and A. Langari, Phys. Rev. A 79, 042319 (2009).
\bibitem{BQL2001} B. Q. Liu, B. Shao, J. G. Li, J. Zou, and L. A. Wu, Phys. Rev. A 83, 052112 (2011).
\bibitem{SHC2018} S. H. Chun, K. W. Shin, H. J. Kim, S. Jung, J. Park, Y. M. Bahk, H. R. Park, J. Kyoung, D. H. Choi, D. S. Kim, G. S. Park, J. F. Mitchell, and K. H. Kim, Phys. Rev. Lett. 120, 027202 (2018).
\bibitem{TAK1983} T.A. Kaplan, Z. Phys. B: Condens. Matter 49, 313 (1983).
\bibitem{LSO1992} L. Shekhtman, O. Entin-Wohlman, and A. Aharony, Phys. Rev. Lett. 69, 836 (1992).
\bibitem{JCM2013} J. Chovan, M. Marder, and N. Papanicolaou, Phys. Rev. B 88, 064421 (2013).
\bibitem{AZS1998} A. Zheludev, S. Maslov, I. Tsukada, I. Zaliznyak, L.P. Regnault, T. Masuda, K. Uchinokura, R. Erwin, and G. Shirane, Phys. Rev. Lett. 81, 5410 (1998).
\bibitem{AZS1999} A. Zheludev, S. Maslov, G. Shirane, I. Tsukada, T. Masuda, K. Uchinokura, I. Zaliznyak, R. Erwin, and L.P. Regnault, Phys. Rev. B 59, 11 432 (1999).
\bibitem{ITJ2000} I. Tsukada, J. Takeya, T. Masuda, and K. Uchinokura, Phys. Rev. B 62, 10 6061(2000).
\bibitem{SMS2012} S. M\"{u}hlbauer, S. Gvasaliya, E. Ressouche, E. Pomjakushina, and A. Zheludev, Phys. Rev. B 86, 024417 (2012).
\bibitem{SMG2017} S. M\"{u}hlbauer, G. Brandl, M. M{\aa}nsson, and M. Garst, Phys. Rev. B 96, 134409 (2017).
\bibitem{HSK2000} H. Shiba, K. Ueda, and O. Sakai, J. Phys. Soc. Jpn. 69, 1493 (2000).
\bibitem{MDL2001} M.D. Lumsden, B.C. Sales, D. Mandrus, S.E. Nagler, and J.R. Thompson, Phys. Rev. Lett. 86, 159 (2001).
\bibitem{IT2003} I. Tsukada, X. F. Sun, Seiki Komiya, A. N. Lavrov, and Yoichi Ando, Phys. Rev. B 67, 224401 (2003).
\bibitem{Ruderman1954} M.A. Ruderman and C. Kittel, Phys. Rev. 96, 99 (1954).
\bibitem{Kasuya1956} T. Kasuya, Prog. Theor. Phys. 16, 45 (1956).
\bibitem{Yosida1957} K. Yosida, Phys. Rev. 106, 893 (1957).
\bibitem{Ren2014} J. Fransson, J. Ren, and J. X. Zhu, Phys. Rev. Lett. 113, 257201 (2014).
\bibitem{TSJ2016} T. Saygun, J. Bylin, H. Hammar, and J. Fransson, Nano Lett. 16, 2824 (2016).
\bibitem{JD2017} J. D. Vasquez Jaramillo and J. Fransson, J. Phys. Chem. C, 121, 27357 (2017).
\bibitem{HH2018} H. Hammar and J. Fransson, arXiv:1806.07092 (2018).
\bibitem{XS2017} X. Shi, H. Yuan, X. Mao, Y. Ma, and H. Q. Zhao, Phys. Rev. A 95, 052332 (2017).
\bibitem{Lee2015} Y. W. Lee and Y. L. Lee, Phys. Rev. B 91, 214431 (2015).
\bibitem{Shiranzaei2017} M. Shiranzaei, H. Cheraghchi and F. Parhizgar, Phys. Rev. B 96, 024413 (2017).
\bibitem{Hasan2010} M. Z. Hasan and C. L. Kane, Rev. Mod. Phys. 82, 3045 (2010).
\bibitem{Qi2011} X. L. Qi and S. C. Zhang, Rev. Mod. Phys. 83, 1057 (2011).
\bibitem{Rashba1960} E. I. Rashba, Sov. Phys. Solid State 2, 1109 (1960).
\bibitem{Bychkov1984} Y. A. Bychkov and E. I. Rashba, J. Phys. C 17, 6039 (1984).
\bibitem{Christopher2016} C. J. Pedder, T. Meng, R. P. Tiwari, and T. L. Schmidt, Phys. Rev. B 94, 245414 (2016).
\bibitem{TS2016} T. Shibuya, H. Matsuura, and M. Ogata, J. Phys. Soc. Jpn 85, 114701 (2016).
\bibitem{VG2005} V. Gritsev, G. Japaridze, M. Pletyukhov, and D. Baeriswy, Phys. Rev. Lett. 94, 137207 (2005).
\bibitem{RM2010} R. M. Lutchyn, J. D. Sau, and S. D. Sarma, Phys. Rev. Lett. 105, 077001 (2010).
\bibitem{YO2010} Y. Oreg, G. Refael, and F. von Oppen, Phys. Rev. Lett. 105, 177002 (2010).
\bibitem{SW1996} N. F. Schwabe, R. J. Elliott, and Ned S. Wingreen, Phys. Rev. B 54,18 (1996).
\end{thebibliography}
\end{document}